\documentclass[twocolumn,superscriptaddress,nofootinbib,floatfix,aps,prb,citeautoscript,longbibliography,10pt]{revtex4-2}

\usepackage[english]{babel}
\usepackage[utf8]{inputenc}
\usepackage[T1]{fontenc}
\usepackage{lineno}
\usepackage{siunitx}
\usepackage{graphicx}
\usepackage{color}
\usepackage{array}
\usepackage{dcolumn}
\usepackage{bm}
\usepackage{multirow}
\usepackage[version=4]{mhchem}
\usepackage{booktabs}
\usepackage{hyperref}
\usepackage{cleveref}
\usepackage{amsmath,amssymb}
\usepackage{placeins}
\usepackage{balance}
\usepackage{orcidlink}
\usepackage{float}

\newcommand{\mevatom}{\ensuremath{\textrm{meV atom}^{-1}}}
\newcommand{\evatom}{\ensuremath{\textrm{eV atom}^{-1}}}
\newcommand{\alex}{\textsc{Alexandria}}
\newcommand{\mbd}{Matbench Discovery}
\newcommand{\matpes}{\textsc{MatPES}}
\newcommand{\wbmmatpes}{WBM-\textsc{MatPES}}
\newcommand{\alexmatpes}{\alex-\textsc{MatPES}}
\newcommand{\tuga}{Tuga-FF}
\newcommand{\ehull}{\ensuremath{E_{\mathrm{hull}}}}
\newcommand{\eps}{\ensuremath{E_{\mathrm{ps}}}}
\newcommand{\daf}{\ensuremath{\mathrm{DAF}}}

\newcommand{\panel}[1]{\noindent{\bfseries (#1)}\par\nointerlineskip\vspace{0.1em}}

\begin{document}

\title{Pressure-transferable universal machine-learning potentials:
  the \wbmmatpes\ benchmark and \tuga}

\author{Pierre-Paul De Breuck}
\affiliation{Research Center Future Energy Materials and Systems of the
  University Alliance Ruhr and Interdisciplinary Centre for Advanced Materials Simulation, Ruhr University Bochum, Universit\"atsstra\ss e 150, D-44801 Bochum, Germany}
 
\author{Thalis H. B. da Silva}
\affiliation{Research Center Future Energy Materials and Systems of the
  University Alliance Ruhr and Interdisciplinary Centre for Advanced Materials Simulation, Ruhr University Bochum, Universit\"atsstra\ss e 150, D-44801 Bochum, Germany}

\author{Antoine Loew}
\affiliation{Research Center Future Energy Materials and Systems of the
  University Alliance Ruhr and Interdisciplinary Centre for Advanced Materials Simulation, Ruhr University Bochum, Universit\"atsstra\ss e 150, D-44801 Bochum, Germany}

\author{Silvana Botti}
\affiliation{Research Center Future Energy Materials and Systems of the
  University Alliance Ruhr and Interdisciplinary Centre for Advanced Materials Simulation, Ruhr University Bochum, Universit\"atsstra\ss e 150, D-44801 Bochum, Germany}
\affiliation{Faculty of Physics and Astronomy, Ruhr University Bochum,
  Universit\"atsstra\ss e 150, D-44801 Bochum, Germany}

\author{Miguel A. L. Marques}
\affiliation{Research Center Future Energy Materials and Systems of the
  University Alliance Ruhr and Interdisciplinary Centre for Advanced Materials Simulation, Ruhr University Bochum, Universit\"atsstra\ss e 150, D-44801 Bochum, Germany}

\begin{abstract}
Universal machine-learning interatomic potentials hold the promise to replace costly \textit{ab initio} calculations in materials science workflows. However, the accuracy and transferability of current models are limited by the availability of training data. We release \wbmmatpes, a dataset of 182 thousand structures recomputed with consistent \matpes\ settings, together with two reference convex hulls of different depths and test sets at six pressures up to 150~GPa. These datasets reproduce a realistic discovery workflow and allow potentials to be ranked systematically at this level of theory, similar to the \mbd\ leaderboard at the Materials Project settings. In this benchmark, five published \matpes\ potentials agree within 12~\mevatom\ at 25~GPa, yet their errors increase dramatically to 0.4–5\evatom\ at 150~GPa. To address this limitation, we develop \tuga, a graph transformer trained on datasets spanning ambient and high-pressure configurations alike. \tuga\ is the most accurate model in the benchmark at ambient pressure, with an energy error of 13.6~\mevatom, and maintains an error of just 21~\mevatom\ at 150~GPa. We believe this benchmark will encourage the community to train higher-accuracy models and generate richer datasets, advancing machine-learned interatomic potentials as a whole.
\end{abstract}

\maketitle

\section{Introduction}

Universal machine-learning interatomic potentials (uMLIPs) are arguably the most consequential application of machine learning to atomistic materials science \cite{Xia2025}. A large number of models is now available~\cite{chen2022m3gnet,deng2023chgnet,batatia2024foundation,merchant2023gnome,park2024sevennet,neumann2024orb,yang2024mattersim,omat24,lysogorskiy2026grace,wood2025uma}, and the underlying architectures and training frameworks have reached a considerable degree of maturity. For a growing set of tasks they can replace density-functional theory (DFT) outright, at a computational cost smaller by several orders of magnitude. 

The factor that currently limits the accuracy of these models is no longer the models themselves, but the data used to train them. The overwhelming majority of the available training data was generated with the Perdew-Burke-Ernzerhof (PBE) functional~\cite{Perdew1996} and inherits, almost without exception, the computational settings adopted by the Materials Project~\cite{jain2013mp}, which were fixed more than a decade ago to meet accuracy requirements considerably looser than those of today. 

Three consequences of this are worth stating explicitly. First, the plane-wave cutoffs and $k$-point densities are modest by present standards, and we roughly estimate that of the order of 3\% of the compounds in these databases are not converged to within 5~\mevatom. The resulting errors are quantitative and, in most applications, tolerable. Second, and more serious, several of the projector augmented-wave (PAW) setups distributed with VASP~\cite{Kresse1996,Kresse1996_2} have been corrected or improved since those settings were fixed. In particular, a number of rare-earth setups keep the $f$ electrons frozen in the core, which leads to an incorrect description of the magnetic moments of many compounds. Third, the Materials Project applies a Hubbard $U$ correction to selected oxides and fluorides~\cite{wang2006ggau,jain2011mixing}. Because this correction is switched on or off as a function of composition, it introduces discontinuities in the potential energy surface, which are difficult to represent with a continuous machine-learning model and are undesirable by construction. An updated and internally consistent set of computational settings is therefore required, and this is precisely what \matpes~\cite{MATPES} provides.

Benchmarks have played an important role in the development of uMLIPs. \mbd~\cite{Riebesell2025} provides a common yardstick against which academic groups and industrial laboratories can measure progress, and has thereby driven a rapid succession of improvements. Its test set is to a large extent based on the Wang-Botti-Marques WBM dataset~\cite{wang2021wbm}, which is computed with Materials Project settings. No equivalent of WBM exists at \matpes\ settings, however, and models trained on \matpes\ cannot be scored against it, since the comparison would measure the mismatch between the two sets of settings rather than the quality of the model. Benchmarks specific to \matpes-trained models have appeared, covering equilibrium properties, molecular-dynamics stability and ionic conductivity~\cite{MATPES}, the physicality of energy-volume curves under extreme hydrostatic strain~\cite{mpaloe2025}, and finite-temperature ensembles~\cite{dynamat2026}.

A benchmark built on relaxations near ambient equilibrium probes only a small region of the potential energy surface, and the training sets of most uMLIPs are themselves dominated by near-equilibrium configurations. This biased sampling is known to soften the predicted surface systematically~\cite{deng2025softening}, and to cost accuracy under compression~\cite{loew2026pressure}. It is therefore not obvious that a ranking established at ambient pressure carries over to the behavior of the same models away from equilibrium, and to our knowledge this has not been tested for \matpes-trained models.

In this work we recompute the WBM dataset with \matpes\ settings and build the corresponding stability benchmark, which includes reference convex hulls derived from the Materials Project and from the \alex\ database~\cite{schmidt2023alexandria,schmidt2024alexandria,cavignac2026alexandria}, also at \matpes\ settings. We then extend the benchmark to six pressures up to 150~GPa, where the compared quantity is the enthalpy. In parallel we present \tuga, an invariant graph-transformer potential trained with data computed at \matpes\ settings. Its architecture is described in \cref{sec:tuga}. The benchmark reported here is itself an experiment with this architecture, and is used to show the impact of the training data on the quality of the force-field.

\section{Results}
\label{sec:results}

\subsection{The \wbmmatpes\ benchmark}
\label{sec:benchmark}

\begin{figure*}[t]
  \centering
  \includegraphics[width=\linewidth]{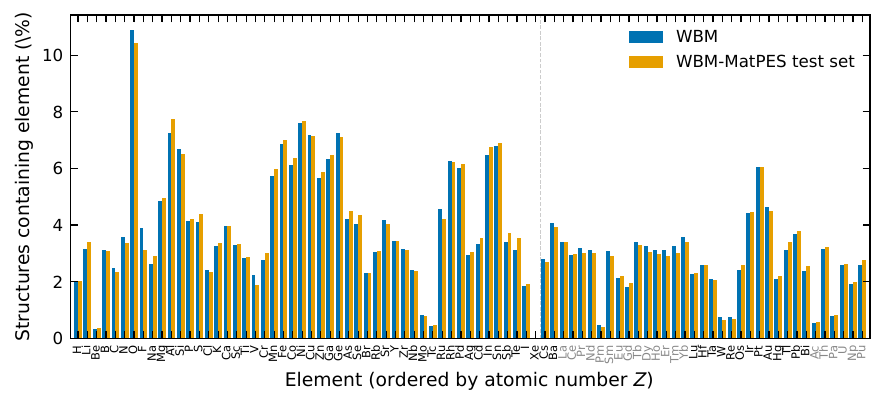}
  \caption{Fraction of structures containing each chemical element, ordered by atomic number $Z$, for the original WBM dataset (blue) and for the \wbmmatpes\ test set (orange). Dashed vertical lines mark the noble-gas atomic numbers, and $f$-block labels are shown in gray.}
  \label{fig:elements}
\end{figure*}

The WBM dataset~\cite{wang2021wbm} was generated by successive elemental substitution into known prototypes. It contains a large number of compounds that were close to, but not in, the Materials Project, and it is the test set of \mbd~\cite{Riebesell2025}. We recomputed 197~thousand WBM structures with \matpes\ settings (see \cref{sec:dft}), of which 191~thousand completed successfully. We then removed compounds that were already included in the reference set, since they could not be considered new discoveries, as well as a small number of entries with unusable unrelaxed geometries, which leaves 182~thousand structures. As in \mbd, all metrics are reported on the subset of unique prototypes, in which only the first of several structures sharing the same AFLOW prototype label and chemical system is kept, so that 179~thousand structures remain. The chemical composition of this test set is essentially indistinguishable from that of the original WBM (\cref{fig:elements}), and the subset retained here therefore inherits the chemical coverage of WBM.

\begin{figure*}[t]
  \centering
  \begin{minipage}[t]{0.42\linewidth}
    \panel{a}
    \includegraphics[width=\linewidth]{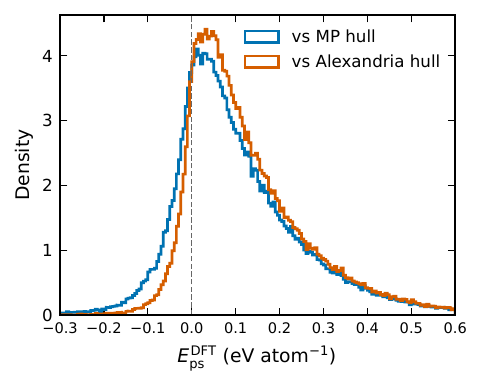}
  \end{minipage}\hspace{0.04\linewidth}%
  \begin{minipage}[t]{0.42\linewidth}
    \panel{b}
    \includegraphics[width=\linewidth]{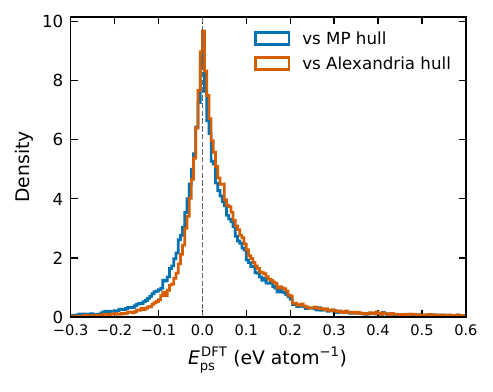}
  \end{minipage}
  \caption{Distribution of the DFT phase separation energy \eps\ evaluated against the Materials-Project-derived hull (blue) and against the \alex\ hull (orange), for (a) the \wbmmatpes\ test set and (b) the Materials Project reference set. The dashed line marks the stability threshold. Against the Materials-Project-derived hull, the reference set is the one that defines the hull.}
  \label{fig:eps}
\end{figure*}

The task follows the initial-structure-to-relaxed-energy protocol of \mbd, in which each model receives the unrelaxed WBM geometry and relaxes atomic positions and cell. Its final energy is converted into a distance to a reference convex hull, which decides whether the compound is predicted to be stable. Two reference hulls are provided, both computed with the same settings and without any energy correction. The first is equivalent to the Materials Project and holds the 89~thousand entries of the Materials Project snapshot used by \mbd\ that we could identify in our database and recompute at \matpes\ settings, of which 31~thousand lie on the hull. The second is the hull of the \alex\ database recomputed at \matpes\ settings, with 166~thousand entries spanning 89 elements, and it is substantially deeper than the Materials-Project-derived one. Stability is assessed with the phase separation energy \eps~\cite{bartel2020}, which measures the distance to the hull formed by all \emph{other} compositions and is therefore well defined even for compounds that are themselves hull vertices (see \cref{sec:protocol}). \Cref{fig:eps} shows the distribution of the DFT phase separation energy of the test set and of the reference set against both hulls. The fraction of the test set that is thermodynamically stable falls from 21.5\% against the Materials-Project-derived hull to 12.5\% against the \alex\ hull, so that roughly four in ten of the apparent discoveries against a Materials-Project-like reference are already beaten by an entry of \alex. This gap is a property of the reference, not of any model, and a benchmark built on a single hull cannot detect it.

\subsection{The \tuga\ model}
\label{sec:models_results}

\tuga\ is an invariant graph neural network that uses the set of interatomic distances and bond angles of the crystal, arranged in an atom graph and in a line graph that carries the three-body angular information. The network predicts a single scalar energy as a sum of atomic contributions. Because this energy is a single continuous function whose derivatives are evaluated exactly, the resulting force field is conservative and equivariant by construction. The backbone is a graph transformer whose attention is conditioned on the bond through feature-wise linear modulation, with element-dependent radial and angular cutoffs. The complete architecture, with every operation of the model, is shown in \cref{fig:architecture} and described in \cref{sec:tuga} and in \cref{app:architecture}. For this benchmark we trained two variants of this model that differ only in their training data. Both were pretrained on the Open Materials 2024 (OMat24) database~\cite{omat24}, and then fine-tuned on \alexmatpes, the set of calculations of the \alex\ database that we performed with \matpes\ settings, which is more than an order of magnitude larger than the released \matpes\ dataset~\cite{MATPES}. \tuga-v1-3d was fine-tuned on its three-dimensional bulk structures, and \tuga-v1-all on the full set, which adds zero-, one- and two-dimensional systems and bulk structures under pressure. A third variant, \tuga-v1-mp, was fine-tuned at Materials Project settings and is used only in \cref{sec:mbd_results}. The WBM test structures and the compounds of the pressure benchmark were excluded from both training sets.

Only models trained, or fine-tuned, on \matpes\ data can be evaluated on this benchmark, since any other uMLIP predicts total energies at a different level of theory, and this restricts the pool of public candidates to five. Four of them, TensorNet, M3GNet, CHGNet and QET, were trained on \matpes-PBE and are distributed through MatGL~\cite{ko2025matgl}. The fifth, MACE-MATPES-PBE-0, was pretrained on OMat24~\cite{omat24} and then fine-tuned on \matpes-PBE. It has therefore been exposed to roughly two orders of magnitude more data than the other four, which should be kept in mind when comparing them.

\subsection{Stability at ambient pressure}
\label{sec:ambient}

\begin{table*}[t]
  \caption{Performance on the \wbmmatpes\ benchmark, evaluated on the
    unique-prototype subset. \ehull\ and \eps\ denote the distance above the
    convex hull and the phase separation energy, respectively. MAE is the mean
    absolute error of the predicted \eps\ in \evatom\ (it is the same for both
    hulls), and RMSD is the root-mean-square displacement with respect to the
    DFT-relaxed geometry, in \AA. The last two columns give the single-point
    (SP) and molecular-dynamics (MD) timings, in ms per structure and in ms per
    step (see \cref{sec:protocol}). \alexmatpes\ denotes the \alex\
    calculations performed with \matpes\ settings, either the 3D bulk
    structures only or the full set (see \cref{app:dataset_breakdown}), and an
    arrow separates pretraining from fine-tuning. Models are ordered by their F1 score on the
    Materials-Project-derived hull.}
  \label{tab:benchmark}
  \small
  \setlength{\tabcolsep}{3pt}
  \begin{ruledtabular}
  \begin{tabular}{llddddddd}
    \multirow{2}{*}{Model} & \multirow{2}{*}{Training data} &
    \multicolumn{2}{c}{MP hull} & \multicolumn{1}{c}{\alex\ hull} &
    \multicolumn{1}{c}{\multirow{2}{*}{MAE}} &
    \multicolumn{1}{c}{\multirow{2}{*}{RMSD}} &
    \multicolumn{1}{c}{\multirow{2}{*}{SP}} &
    \multicolumn{1}{c}{\multirow{2}{*}{MD}} \\
    \cline{3-4}\cline{5-5}
    & & \multicolumn{1}{c}{F1$_{\ehull}$} & \multicolumn{1}{c}{F1$_{\eps}$}
      & \multicolumn{1}{c}{F1$_{\eps}$} & & & & \\
    \hline
    \tuga-v1-all  & OMat24 $\to$ \alexmatpes\ (all) & 0.919 & 0.935 & 0.899 & 0.0136 & 0.055 &  9.0 & 407.8 \\
    \tuga-v1-3d   & OMat24 $\to$ \alexmatpes\ (3D)  & 0.918 & 0.933 & 0.894 & 0.0136 & 0.056 &  8.5 & 408.7 \\
    MACE-MATPES-0 & OMat24 $\to$ \matpes        & 0.754 & 0.793 & 0.694 & 0.0372 & 0.068 & 12.4 & 113.5 \\
    TensorNet     & \matpes                     & 0.730 & 0.773 & 0.667 & 0.0427 & 0.075 & 11.2 &  44.7 \\
    CHGNet        & \matpes                     & 0.686 & 0.733 & 0.615 & 0.0473 & 0.085 & 14.4 & 108.6 \\
    M3GNet        & \matpes                     & 0.676 & 0.723 & 0.602 & 0.0499 & 0.092 & 11.2 &  83.6 \\
    QET           & \matpes                     & 0.599 & 0.648 & 0.507 & 0.0619 & 0.080 & 12.6 &  25.8 \\
  \end{tabular}
  \end{ruledtabular}
\end{table*}

\Cref{tab:benchmark} collects the performance of the seven models on the \wbmmatpes\ benchmark. Among the models trained on the previously released \matpes\ dataset alone, TensorNet is the most accurate, followed by CHGNet, M3GNet and QET. This ordering is the same for the two reference hulls and for the two stability criteria, and it also follows the mean absolute error of the predicted phase separation energy, which ranges from 43~\mevatom\ for TensorNet to 62~\mevatom\ for QET.

MACE-MATPES-0 outperforms all these models, with $\mathrm{F1}_{\eps}=0.793$ against the Materials-Project-derived hull, compared with 0.773 for TensorNet. Because this model was pretrained on OMat24 before being fine-tuned on \matpes, it has seen roughly two orders of magnitude more data than the remainder of that group. The improvement that this yields, 0.02 in F1 and 5~\mevatom\ in the mean absolute error, is measurable but small in comparison with the size of the OMat24. Since OMat24 was computed with Materials Project settings, this small gain suggests that the accuracy attainable at a given set of settings depends mainly on the amount of data computed with those same settings, and much less on the total amount of data seen by the model.

The two \tuga\ models are substantially more accurate than the rest of the set. Against the Materials-Project-derived hull they reach $\mathrm{F1}_{\eps}=0.935$ and 0.933, against 0.793 for the best of the other models, and their mean absolute error of 13.6~\mevatom\ is smaller than that of MACE-MATPES-0 by a factor of 2.7. They are also the only models for which precision and recall are balanced, 0.94 and 0.93 against the Materials-Project-derived hull, whereas every other model trades one against the other and most lose recall disproportionately. Both variants were trained on \alexmatpes, a body of calculations performed with the settings of \cref{sec:dft} that is considerably larger than \matpes\ itself (see \cref{sec:training}). At ambient pressure the two \tuga\ variants are indistinguishable. Their mean absolute errors differ by only $0.004 \pm 0.035$~\mevatom, and the more general model is the more accurate of the two on 49.6\% of these structures. Adding training data from other dimensionalities and from finite pressure therefore has no measurable cost for bulk structures at ambient conditions, a point to which we return in \cref{sec:pressure}.

The root-mean-square displacement with respect to the DFT-relaxed geometries follows nearly the same ordering as the energetic metrics, from 0.055~\AA\ for \tuga-v1-all to 0.092~\AA\ for M3GNet. More than 98\% of the relaxations converge for every model, and the structures that do not converge are excluded from the accuracy metrics.

The computational cost is measured in two ways, which give very different pictures. In single-point evaluations on the small cells of the test set, all models lie within a factor of two, from 8.5~ms per structure for \tuga-v1-3d to 14.4~ms for CHGNet, and the \tuga\ models are the fastest of the set. In a molecular-dynamics step of a 1000-atom cell, the spread is much larger, from 25.8~ms for QET to 409~ms for \tuga-v1-3d, and the \tuga\ models are the slowest, about 3.6 times slower than MACE-MATPES-0. The difference comes from the size of the cells, since for the small cells of the test set the cost is dominated by the fixed overhead of each evaluation, whereas for the 1000-atom cell it is dominated by the arithmetic of the network itself. The cost of \tuga\ is therefore small for applications built on relaxations of small cells, but it matters for extended finite-temperature sampling of large cells. \tuga\ can also be run with reduced precision, which leaves the single-point cost unchanged and lowers the cost of a molecular-dynamics step by a factor of three, but degrades the predictions under pressure, so all results reported here were obtained at full precision. The memory footprint and evaluation time of \tuga\ also grow rapidly for systems of several hundred atoms, and improving this scaling will be the subject of future versions.

\subsection{Dependence on the reference hull}
\label{sec:hulldep}

All models perform worse against the \alex\ hull, which is the deeper of the two hulls, and their $\mathrm{F1}_{\eps}$ drops by between 0.036 for \tuga-v1-all and 0.141 for QET. The ranking of the models is nevertheless unchanged, which indicates that the conclusions drawn from this benchmark do not depend on the choice of reference. The magnitude of the reduction is informative: A model that identifies stable compounds with a precision of 0.84 against a Materials-Project-like reference, as MACE-MATPES-0 does, achieves only 0.75 when the same predictions are confronted with the more complete \alex\ database. This loss is a consequence of  the reference as well as the model, and only the comparison between two references separates the two contributions. The models that are more accurate in absolute terms also lose less, and the coefficient of determination of the \tuga\ models is essentially unchanged between the two hulls, 0.965 and 0.953.

\subsection{\mbd\ benchmark}
\label{sec:mbd_results}

To establish direct comparability with the wider literature on the standard Materials Project reference framework, we evaluated \tuga-v1-mp on the canonical \mbd\ benchmark~\cite{Riebesell2025}. For this comparison, \tuga-v1-mp was fine-tuned on non-\matpes\ data, combining the Materials Project trajectory (MPtrj)~\cite{deng2023chgnet} and the \alex\ database computed with standard Materials Project settings~\cite{schmidt2022alexandria}. It is evaluated on the 179~thousand unique prototypes of the original WBM dataset~\cite{wang2021wbm} against the standard Materials Project reference hull.

\begin{table}[t]
  \caption{Performance of \tuga-v1-mp on \mbd\ benchmark
    (unique-prototype subset of WBM). MAE and RMSE of the formation energy are in \evatom, RMSD is in \AA, and $\kappa_{\mathrm{SRME}}$
    is the symmetric relative error of the lattice thermal conductivity.}
  \label{tab:mbd}
  \small
  \begin{ruledtabular}
  \begin{tabular}{cccccc}
    F1 & DAF & MAE & RMSE & RMSD & $\kappa_{\mathrm{SRME}}$ \\
    \hline
    0.910 & 6.006 & 0.020 & 0.067 & 0.0643 & 0.3808 \\
  \end{tabular}
  \end{ruledtabular}
\end{table}

\Cref{tab:mbd} summarizes the main scores. On thermodynamic stability and structure prediction the metrics are excellent, with an F1 score of 0.910, a discovery acceleration factor ($\mathrm{DAF}$) of 6.01, an energy MAE of 0.020~\evatom\ (20~\mevatom), and an RMSD of 0.0643~\AA.

In contrast to the energetic and geometric metrics, the error on the lattice thermal conductivity ($\kappa_{\mathrm{SRME}} = 0.3808$) is only average across the leaderboard. A closer inspection of the force constants reveals that, while second-order force constants (FC2) and harmonic phonon dispersions are reproduced well, third-order force constants (FC3) deteriorate. Because the model is trained only on energies, forces, and stresses without higher-order supervision, its description of anharmonic properties is an emergent behavior. Most models that achieve low $\kappa_{\mathrm{SRME}}$ on this benchmark employ equivariant message-passing representations or atomic cluster expansions, and improving FC3 will be a topic of investigation for future versions of the model.

\subsection{Behavior under pressure}
\label{sec:pressure}

The benchmark described so far probes the potential energy surface only in the vicinity of the ambient-pressure equilibrium. We therefore extended it to six pressures, 25, 50, 75, 100, 125 and 150~GPa, using the structures of Ref.~\cite{loew2026pressure} recomputed with \matpes\ settings (see \cref{sec:pressure_methods}). The task is the same as at ambient pressure, except that the relaxed quantity is the enthalpy $H=E+PV$ and no convex hull is involved. Between \num{7473} and \num{8173} structures are available per pressure, 48~thousand in total, none of which was used to train the \tuga\ models.

\begin{figure*}[t]
  \centering
  \begin{minipage}[t]{0.32\linewidth}
    \panel{a}
    \includegraphics[width=\linewidth]{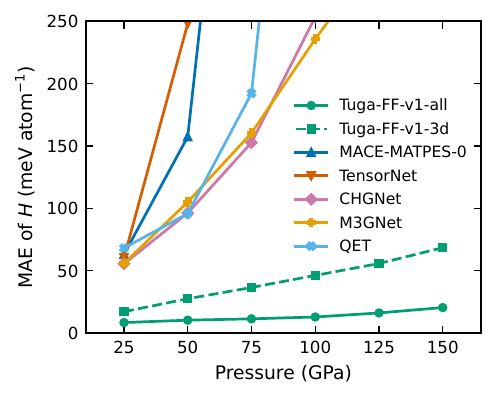}
  \end{minipage}\hfill
  \begin{minipage}[t]{0.32\linewidth}
    \panel{b}
    \includegraphics[width=\linewidth,trim=-11.3 0 0 0]{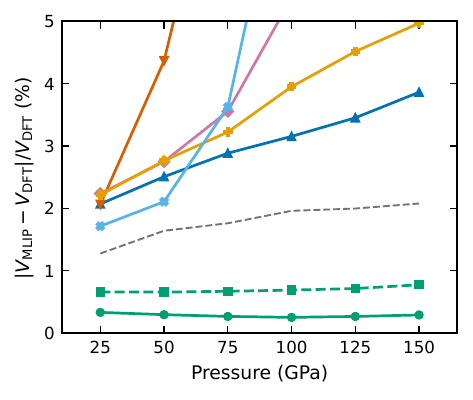}
  \end{minipage}\hfill
  \begin{minipage}[t]{0.32\linewidth}
    \panel{c}
    \includegraphics[width=\linewidth,trim=-2.1 -0.1 0 -3.0]{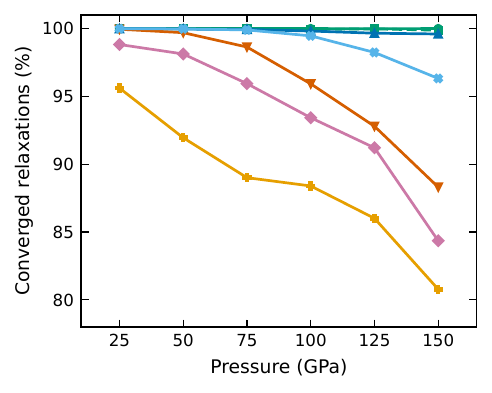}
  \end{minipage}
  \caption{Pressure benchmark. (a) Mean absolute error of the enthalpy per atom, (b) mean absolute relative error of the volume per atom, and (c) fraction of relaxations that reach the force threshold within the step budget, as functions of the applied pressure. In (a) and (b) the averages run over the converged relaxations only. The dashed gray line in (b) is the error of the unrelaxed input, that is, of returning the structure given to the model without modifying it. The vertical axes of (a) and (b) are truncated at 250~\mevatom\ and at 5\%, and that of (c) starts at 78\%. Curves that leave a frame continue towards the values listed in \cref{tab:pressure}. The two \tuga\ variants are drawn in the same color and distinguished by line style, since they form the controlled pair. The legend of (a) applies to all three panels.}
  \label{fig:pressure}
\end{figure*}

\begin{table*}[t]
  \caption{Pressure benchmark at 150~GPa. Conv.\ is the fraction of
    relaxations that converged within the step budget. MAE and $|\Delta V|/V$
    are the mean absolute error of the enthalpy per atom and the mean absolute
    relative error of the volume, both over the converged relaxations, and
    $P\,\Delta V$ and $\Delta E$ are the two terms of the corresponding mean
    signed error of the enthalpy, $\Delta H=\Delta E+P\,\Delta V$. Median and
    ``$>1$~eV'' are the median absolute error of the enthalpy and the fraction
    of relaxations in error by more than 1~\evatom, both over all relaxations
    that returned an energy, converged or not. RMSD is the root-mean-square
    displacement with respect to the DFT geometry over the converged
    relaxations, in \AA. For comparison, the volume error of the unrelaxed
    input is 2.08\%. Energies in \mevatom.}
  \label{tab:pressure}
  \small
  \begin{ruledtabular}
  \begin{tabular}{ldddddddd}
    Model & \multicolumn{1}{c}{Conv.\ (\%)} & \multicolumn{1}{c}{MAE} &
    \multicolumn{1}{c}{$P\,\Delta V$} & \multicolumn{1}{c}{$\Delta E$} &
    \multicolumn{1}{c}{Median} & \multicolumn{1}{c}{$>1$~eV (\%)} &
    \multicolumn{1}{c}{$|\Delta V|/V$ (\%)} & \multicolumn{1}{c}{RMSD} \\
    \hline
    \tuga-v1-all  & 100.0 &   21 &    -5 &    -5 &    9 &  0.2 &  0.29 & 0.015 \\
    \tuga-v1-3d   &  99.9 &   69 &   -29 &    -8 &   43 &  0.4 &  0.78 & 0.028 \\
    MACE-MATPES-0 &  99.6 & 1937 &  -125 & -1700 &  202 &  5.0 &  3.86 & 0.105 \\
    TensorNet     &  88.3 & 4976 & -3519 & -1407 & 6392 & 57.0 & 31.30 & 0.495 \\
    CHGNet        &  84.3 &  570 &  -914 &   384 &  334 & 19.8 &  8.18 & 0.081 \\
    M3GNet        &  80.8 &  381 &  -444 &   138 &  261 & 15.0 &  4.97 & 0.071 \\
    QET           &  96.3 & 3136 & -2776 &  -242 &  356 & 39.9 & 25.41 & 0.376 \\
  \end{tabular}
  \end{ruledtabular}
\end{table*}

\Cref{fig:pressure} shows the error of the enthalpy, the error of the volume and the fraction of converged relaxations as a function of pressure, and \cref{tab:pressure} collects the corresponding numbers at 150~GPa. All models degrade under compression, but by amounts that differ by more than an order of magnitude.

The five previously published models are nearly indistinguishable at the lowest pressure of the grid, 25~GPa, where their errors on the enthalpy lie between 56 and 68~\mevatom. At 150~GPa the same five models span 381 to 4976~\mevatom, a factor of thirteen. Since four of these models were trained on the same data at the same settings, and all five were evaluated on the same structures, this spread is a property of the architectures and of their behavior outside the region the training set covers, not of the data. TensorNet is the most accurate of the \matpes-only models at ambient pressure and the least accurate at 150~GPa, M3GNet moves from the second-worst position to the best, and MACE-MATPES-0, the most accurate of the five at ambient pressure, ranks only third under strong compression.

The failure is of two distinct kinds, which the decomposition of the enthalpy error in \cref{tab:pressure} separates. TensorNet and QET have volume errors of 31 and 25\% at 150~GPa, and the corresponding $P\,\Delta V$ terms, $-3519$ and $-2776$~\mevatom, account for most of their error. Their repulsive wall is too soft, so the cell collapses under the applied pressure. MACE-MATPES-0 behaves in the opposite way. Its volume error at the same pressure is 3.9\%, the smallest of the five, and its $P\,\Delta V$ term is only $-125$~\mevatom, so essentially the whole of its error, $-1700$~\mevatom, sits in the energy assigned to a geometry that is close to the correct one.

The volume errors in \cref{fig:pressure}b should be read against the dashed line, which is the error of the structure handed to the model. Each DFT relaxation at a given pressure was itself started from a geometry relaxed at that pressure with GRACE~\cite{lysogorskiy2026grace}, so the input is already within 1.3 to 2.1\% of the reference volume, and a model that returned it unchanged would achieve that error. None of the five previously released models reaches it at any pressure of the grid. Their relaxations move the cell away from the DFT answer rather than towards it, by a factor between 1.3 and 1.8 at 25~GPa and by up to 15 at 150~GPa, and for these models the geometric part of the benchmark therefore measures the damage done by the relaxation rather than its quality.

The two \tuga\ models, \tuga-v1-all and \tuga-v1-3d, behave differently on every count. Their mean error on the enthalpy grows from 8 and 17~\mevatom\ at 25~GPa to 21 and 69~\mevatom\ at 150~GPa, respectively, so that at the highest pressure of the grid \tuga-v1-all is still about three times more accurate than any other model at the lowest. For \tuga-v1-all, the decomposition shows no dominant contribution of either kind, with $P\,\Delta V$ and energy terms both equal to $-5$~\mevatom\ at 150~GPa. The two variants are the only models whose volume error lies below that of the unrelaxed input, 0.29 and 0.78\% against 2.08\% at 150~GPa, and the only ones for which that error does not grow appreciably with pressure. Their median error at 150~GPa, 9 and 43~\mevatom, is smaller than that of any other model by a factor of at least 22 and 5, respectively, and the fraction of relaxations in error by more than 1~\evatom\ is 0.2 and 0.4\%, against 5\% for MACE-MATPES-0 and 15 to 57\% for the models trained on \matpes\ alone. Together with MACE-MATPES-0, they are also the only models that relax essentially every structure to the force threshold at every pressure of the grid.

The comparison between the two \tuga\ variants isolates the effect of the training data. They share the architecture and the pretraining, and the fine-tuning set of \tuga-v1-all is a superset of that of \tuga-v1-3d, adding low-dimensional systems, the released \matpes\ dataset and bulk structures under pressure. Since the two are indistinguishable at ambient pressure, the additional data brings no general advantage, and the difference that appears under compression is attributable to the compressed structures it contains. The variant that has seen compressed environments is more accurate at every pressure and on every quantity we report, and the margin widens with compression. Its mean error is smaller by a factor of 2.0 at 25~GPa and of 3.3 at 150~GPa, its median error by a factor of 2.4 and 4.8, and its volume error by a factor of two throughout. This is the expected result, and it reproduces at \matpes\ settings the conclusion of Ref.~\cite{loew2026pressure} that the degradation of uMLIPs under compression originates in the coverage of the training set. The improvement also extends to the tail of the distribution, although the tail is small for both variants, with only 16 relaxations of \tuga-v1-all and 30 of \tuga-v1-3d, out of \num{7473}, in error by more than 1~\evatom\ at 150~GPa.

Two remarks accompany these results. The first concerns the convergence of the relaxations, shown in \cref{fig:pressure}c. Both \tuga\ variants and MACE-MATPES-0 reach the force threshold for essentially every structure at every pressure, and QET for more than 96\%, whereas for TensorNet, CHGNet and M3GNet the converged fraction falls with compression, to between 81 and 88\% at 150~GPa. At that pressure, 308 relaxations with CHGNet and 154 with M3GNet terminated with an error and returned no energy, and these are counted as unconverged. For TensorNet, QET and MACE-MATPES-0, the unconverged relaxations are collapsed cells, with mean volume errors of about 50 to 80\% and enthalpy errors of several \evatom. The second remark is that the mean is not a robust summary of these distributions. At 150~GPa the median error of MACE-MATPES-0 is 202~\mevatom\ against a mean of 1937, the difference being carried by a few structures on which the relaxation diverges numerically, while more than half of the structures relaxed with TensorNet are in error by more than 1~\evatom\ and its median error is 6.4~\evatom. We report the mean for consistency with the ambient-pressure benchmark, but the median and the number of catastrophic failures in \cref{tab:pressure} are the more informative quantities in this regime.

\section{Discussion}
\label{sec:discussion}

This work had two purposes, to provide a reference against which potentials trained on \matpes\ can be scored consistently, and to use this reference to ask what governs their accuracy and their transferability. The evaluation of seven models leads to four observations.

The first is that the ranking of the models does not depend on the reference against which they are measured, although their absolute accuracy does. The \alex\ hull contains 166~thousand entries against the 31~thousand of the Materials-Project-derived one, and the fraction of the test set that is thermodynamically stable falls from 21.5\% to 12.5\% when it is used. All models lose accuracy against this deeper reference, by between 0.04 and 0.14 in F1, but they lose it in the same order. This is a useful property for a benchmark, because the conclusions do not hinge on a choice that is to some extent arbitrary. At the same time, it sets a limit on what a single-hull benchmark can tell us, since the absolute precision of a model is a statement about the reference as much as about the model.

The second observation concerns the training data. MACE-MATPES-0 was pretrained on OMat24 and then fine-tuned on \matpes, so it has seen around two orders of magnitude more data than the models trained on the \matpes\ dataset alone. The gain is real but small, amounting to 0.02 in F1 and 5~\mevatom\ in the mean absolute error with respect to TensorNet, the best model of that group. The \tuga\ models, in contrast, improve on the same reference by 0.14 in F1 and by a factor of 2.7 in the mean absolute error, and they were trained on \alexmatpes, a much larger body of calculations performed with the settings used here. What distinguishes the two cases is therefore not the quantity of data but the settings with which it was computed. OMat24 follows Materials Project settings, and a large corpus at those settings transfers only weakly to a target defined by different ones. The same architecture makes the point directly. \tuga-v1-mp, fine-tuned at Materials Project settings, reaches 20~\mevatom\ on \mbd, where the lowest reported errors are 18~\mevatom, while the \matpes-trained variants reach 13.6~\mevatom\ on the same WBM structures scored against our own reference. Comparing each model with the level of theory it was trained on is the only comparison that carries information, and it shows that the target limits the accuracy attainable, since settings that are discontinuous in composition impose a floor that training cannot remove. To our knowledge 13.6~\mevatom\ is the lowest error reported for a universal potential on WBM at the PBE level, which makes \tuga\ the most accurate predictor of PBE ground-state energies available, a property of the architecture and the training data taken together. We also note that the two \tuga\ variants are indistinguishable at ambient pressure, although one of them also saw low-dimensional systems and structures under pressure. This is unexpected, since multi-task training usually costs accuracy on each individual task.

The third observation is the one that we consider most consequential, namely that agreement at ambient pressure does not imply agreement away from it. The five previously released models were all trained, or fine-tuned, on \matpes\ and lie within 12~\mevatom\ of each other at 25~GPa, yet they span a factor of thirteen at 150~GPa. None of them improves, at any pressure of the grid, on the volume that would be obtained by returning the input structure unchanged, and between 5 and 57\% of their relaxations at 150~GPa are in error by more than 1~\evatom. This is the expected consequence of evaluating a model outside the region its training set covers. It reproduces at these settings the behavior reported in Ref.~\cite{loew2026pressure} for models trained with Materials Project settings, as well as the softening documented in Ref.~\cite{deng2025softening}. The spread among architectures is a separate matter, because architectures that are indistinguishable within their training domain need not remain so outside it. The failure also takes different forms. For TensorNet and QET the cell collapses, and the error comes mainly from the volume term $P\,\Delta V$, whereas for MACE-MATPES-0 the geometry remains close to the reference and the error comes mainly from the energy itself. A single accuracy metric cannot tell these two cases apart.

The fourth observation is that the deficiency is one of coverage rather than of architecture. The two \tuga\ variants share an architecture and a pretraining, and the training set of \tuga-v1-all is a superset of that of \tuga-v1-3d which adds structures under pressure. Since the two are indistinguishable at ambient pressure, that superset brings no general advantage, and the difference that appears under compression follows from the compressed structures it adds. This variant is more accurate at every pressure and on every quantity we report, by a margin that widens with compression from a factor of two at 25~GPa to a factor of three at 150~GPa, and its mean error at the highest pressure of the grid, 21~\mevatom, is about three times smaller than that of any other model at the lowest. Both variants relax essentially every structure to the force threshold at every pressure, and their catastrophic failures are rare, 16 and 30 of \num{7473} relaxations at 150~GPa, so the additional data improves the tail of the distribution as well as its bulk. Covering the relevant region of configuration space is therefore enough to remove the degradation almost entirely, without any change to the model.

This further confirms that a potential should not be assumed to transfer to a region of configuration space that its training set does not cover, and that a ranking established in one region carries no guarantee for another. Benchmarks that probe a single region are therefore not sufficient on their own, however carefully they are built.

Finally, a remark on scope and limitations. The benchmark measures the prediction of thermodynamic stability from a relaxation, which is one task among several, and accuracy on it does not guarantee accuracy on phonons, on transport properties, or in long molecular-dynamics trajectories. As the \mbd\ results for \tuga-v1-mp illustrate, low errors on energies and relaxed geometries do not ensure accurate lattice thermal conductivities when third-order force constants deteriorate. The \matpes\ benchmarks covering those aspects~\cite{MATPES,mpaloe2025,dynamat2026} are complementary to this one, and we would discourage the use of any single number as a summary of the quality of a potential.

\section{Methods}
\subsection{\tuga\ architecture and training}
\label{sec:tuga}

\tuga\ is an invariant graph transformer built exclusively from interatomic distances and bond angles, avoiding spherical harmonics and tensor products. The total potential energy is evaluated as an extensive sum of atomic contributions,
\begin{equation}
  E = \sum_i \varepsilon_i,
  \qquad
  \varepsilon_i = \mathrm{MLP}_E\big(\bm h_i^L\big) + \epsilon_{Z_i},
  \label{eq:energy_readout}
\end{equation}
where $\bm h_i^L$ denotes the final node vector of atom $i$ after $L=4$ interaction layers, and $\epsilon_{Z_i}$ is a species-dependent elemental reference energy.

Because the potential energy surface $E$ is four times continuously differentiable ($C^4$) with respect to atomic coordinates and strain, atomic forces $\bm F_i$ and the Cauchy stress tensor $\bm\sigma$ are obtained analytically via automatic differentiation:
\begin{equation}
  \bm F_i = -\frac{\partial E}{\partial \bm r_i},
  \qquad
  \bm\sigma = \frac{1}{V}\frac{\partial E}{\partial \bm\varepsilon}\Big|_{\bm\varepsilon=0},
  \label{eq:forces_stress}
\end{equation}
where $V$ is the cell volume and $\bm\varepsilon$ is a virtual strain tensor. By evaluating forces as exact gradients of a single scalar energy, the potential is strictly conservative and equivariant under rotations, reflections, and translations by construction, without requiring equivariant tensor features. An architectural overview is given in \cref{fig:architecture}, and complete hyperparameters are listed in \cref{app:architecture}.

Compared with standard message-passing graph neural networks, the design of \tuga\ is guided by four distinct physical and algorithmic principles.

\textbf{Decoupled element-dependent cutoffs.} Rather than applying a uniform spatial cutoff to all elements, \tuga\ assigns each of the 89 elements its own cutoff spheres, a bond radius $r_c(Z)$ for two-body interactions and a triplet radius $r_c^\tau(Z)$ for three-body interactions. For a pair $(i, j)$ the cutoff is the arithmetic mean $r_c^{ij} = \tfrac{1}{2}(r_c(Z_i)+r_c(Z_j))$, while for a triplet centered on atom $i$ the leg cutoff is $r_c^\tau(Z_i)$. Every interatomic distance $r$ is windowed by a smooth $C^4$ polynomial envelope $\chi(r)$ that decays to zero across a boundary width $w = 1.0$~\AA, which ensures strictly continuous forces ($C^3$) and vibrational Hessians ($C^2$). Its exact algebraic formulation is given in \cref{app:cutoff_poly}.

\textbf{Collinear singularity avoidance.} Three-body angular features are parameterized directly through the bond-angle cosine, $\cos\theta_{jik} = (\bm r_{ij}\cdot\bm r_{ik})/(r_{ij} r_{ik})$. Because the derivative $\frac{d}{d\bm r}\arccos(\cos\theta)$ diverges as $\theta\to 0$ or $\pi$, formulating the three-body channel in $\cos\theta$ avoids collinear force singularities without artificial angular cutoffs.

\textbf{Custom FiLM-modulated graph attention with physical continuity.} Unlike standard GNNs that use isotropic message passing or vanilla self-attention, \tuga\ alternates between line-graph and atom-graph attention (\cref{fig:architecture} b,c). In both graphs, angle and edge vectors dynamically modulate the attention keys and values through feature-wise linear modulation (FiLM) across $H=8$ heads $\eta$, and the attention scores are conditioned on the angle basis $\bm\psi(\theta)$ and the radial Bessel basis $\bm\phi(r)$ through learned gates and biases. Crucially, standard softmax attention normalizes weights to unity ($\sum_j \alpha_{ij} = 1$), which causes an isolated neighbor crossing the cutoff boundary to jump discontinuously from $\alpha_{ij} = 1$ to 0. \tuga\ eliminates this boundary discontinuity by multiplying each message $\bm m_{ij}^l$ directly by the smooth envelope $\chi(r_{ij})$ (\cref{fig:architecture}c, \cref{app:film_attn}). All atomic interactions thus vanish smoothly as neighbors cross the cutoff sphere.

\textbf{Memory-efficient symmetric triplet pooling.} To pass representations between atoms, bonds, and angles, \tuga\ uses a symmetric mixer (SymUpdate). Standard line-graph networks materialize full unrolled triplet tensors of shape $(N_{\mathrm{triplets}}, 2d)$, which causes severe memory bottlenecks. SymUpdate encodes source states once on the nodes before pair indexing, and fuses them via normalized sum $\bm s = (\tilde{\bm u} + \tilde{\bm w})/\sqrt 2$ and Hadamard product $\bm p = \tilde{\bm u} \odot \tilde{\bm w}$ pooling. This enforces exact permutation invariance with respect to flanking bonds while reducing the peak memory footprint several-fold (\cref{app:sym_update}).

\phantomsection\label{sec:training}%
Atoms enter the network via a hybrid embedding $\bm u_i = \big[\bm P_{Z_i} \Vert \bm L_{Z_i}\big]$, which concatenates a 98-dimensional physical similarity prior $\bm P_{Z}$~\cite{cerqueira2026representation} with a 128-dimensional learned vector $\bm L_{Z}$. Attention blocks are coupled with SwiGLU feed-forward layers~\cite{shazeer2020glu} and RMSNorm~\cite{zhang2019rmsnorm} in a pre-norm residual topology~\cite{xiong2020prenorm}.

The potential is trained on total energies, atomic forces, and Cauchy stresses using a multi-task Huber loss~\cite{huber1964}. The exact loss formulation, weights, and thresholds are given in \cref{app:loss} and \cref{tab:arch_hparams}. Optimization was performed with Adam~\cite{kingma2015adam} using a one-cycle learning-rate schedule~\cite{smith2019onecycle} with 5\% linear warm-up, gradient norm clipping at 0.5, and dynamic batch packing to fill memory.

All models were first pretrained on 89.5~million configurations from OMat24~\cite{omat24} ($\approx 1400$ GPU-hours on NVIDIA H100 GPUs). Three specialized models were then fine-tuned: (i)~\tuga-v1-mp on 12~million configurations combining MPtrj~\cite{deng2023chgnet} and \alex~\cite{schmidt2022alexandria} at standard Materials Project settings, (ii)~\tuga-v1-3d on the 7~million 3D bulk frames of \alexmatpes, the set of \alex\ calculations performed with \matpes\ settings, and (iii)~\tuga-v1-all on the 21.4~million frames of the full \alexmatpes\ set, encompassing 0D to 3D systems, high-pressure phases up to 150~GPa, and the released \matpes\ dataset~\cite{MATPES}. WBM test structures and pressure benchmark compounds were strictly withheld from all training corpora. A full breakdown of datasets, weights, and hardware budgets is given in \cref{app:architecture}.

\subsection{Computational details}
\label{sec:dft}

All DFT calculations were carried out with the Vienna \textit{Ab initio} Simulation Package (VASP)~\cite{Kresse1996,Kresse1996_2}, following the \matpes\ settings~\cite{MATPES}. Exchange and correlation were treated with the PBE functional~\cite{Perdew1996}, and the electron-ion interaction was described by PAW pseudopotentials~\cite{Blchl1994,Kresse1999} with a plane-wave cutoff of 680~eV. The Brillouin zone was sampled with $\Gamma$-centered $k$-point grids of spacing 0.22~\AA$^{-1}$, and a Gaussian smearing of width $\sigma=0.05$~eV was applied. All calculations were spin polarized, with convergence thresholds of $10^{-7}$~eV for the total energy and 5~meV~\AA$^{-1}$ for the forces. The smearing correction proposed as part of \matpes\ was not applied, as its implementation was incompatible with the high-throughput workflow used here.

We note that the PAW setups prescribed by the \matpes\ settings are more recent than those employed for the Materials Project~\cite{jain2013mp} and for the ambient-pressure \alex\ database~\cite{schmidt2023alexandria,schmidt2024alexandria,cavignac2026alexandria}, and were selected for their improved accuracy. Energies obtained with these settings are therefore not interchangeable with those of either database, which is the reason why a dedicated test set and dedicated reference hulls are required.

\subsection{Construction of the test set}
\label{sec:testset}

A dataset of 197~thousand WBM structures~\cite{wang2021wbm} was recomputed with the settings of \cref{sec:dft}, of which 191~thousand completed successfully within our computational budget. From these we removed the \num{7488} structures that also belong to the Materials Project reference set described in \cref{sec:hulls}. We also removed \num{1181} entries whose unrelaxed geometry was unavailable, or described a supercell of the relaxed one. The resulting test set contains 182~thousand structures.

Following the convention of \mbd~\cite{Riebesell2025}, all metrics are reported on the subset of unique prototypes, obtained by retaining only the first occurrence of each combination of AFLOW prototype label and chemical system. This subset comprises 179~thousand structures, or 98.1\% of the test set. The fraction removed is considerably smaller than in \mbd, where the same criterion discards 16\% of WBM, because duplicate structures are already rejected when entries are inserted into our database. The choice moves the F1 score by less than 0.001.

\subsection{Reference convex hulls}
\label{sec:hulls}

Two independent reference convex hulls are used, both computed with the settings of \cref{sec:dft}.

The first is derived from the Materials Project. Starting from the snapshot of 2023-02-07 used by \mbd~\cite{Riebesell2025}, which contains 155~thousand entries, each structure was identified in our database of calculations performed with Materials Project settings, and the corresponding entry recomputed with \matpes\ settings was retained. Identification was performed at the original settings rather than at \matpes\ settings, because relaxation with the latter can displace a structure far enough that it no longer matches its parent. The structure matching used the \texttt{StructureMatcher} of \texttt{pymatgen}~\cite{ong2013pymatgen} with primitive-cell reduction, without volume rescaling, and with an element-only comparator. For each Materials Project entry we enumerated all candidates of the same reduced formula and kept the one with the smallest root-mean-square distance. Keeping the first match instead would preferentially select the lowest-energy polymorph among several that fit, and would bias the hull downwards. Tolerances of $(\ell,s,\theta)=(0.10,0.15,2.5^{\circ})$ were applied first, falling back to the \texttt{pymatgen} defaults of $(0.20,0.30,5^{\circ})$ only when no match was found. Matches accepted at the looser tolerance with a root-mean-square distance above 0.15 were discarded. In this way 125~thousand Materials Project entries were matched. After the tolerance cut, and after removing entries claimed by more than one Materials Project material, 117~thousand distinct structures remained. Of these, 89~thousand have a converged counterpart at \matpes\ settings. The remainder did not converge within our computational budget, mostly because of the size of their unit cells. Of the 89~thousand entries, 31~thousand lie on the resulting hull. All elemental references required by the test set are present in this set, so that no endpoint had to be added by hand.

The second hull is the \alex\ hull, taken as the full snapshot of the database that has been recomputed with \matpes\ settings, comprising 166~thousand entries spanning 89 chemical elements. It is substantially deeper than the Materials-Project-derived hull, and the comparison between the two quantifies how much of the apparent performance of a model depends on the completeness of the reference.

No energy corrections were applied at any stage. The MP2020 compatibility scheme~\cite{wang2021mp2020} exists to reconcile the mixture of GGA and GGA$+U$ calculations present in the Materials Project~\cite{jain2011mixing}, and is not applicable to the internally consistent, $U$-free energies used here.

\subsection{Evaluation protocol and metrics}
\label{sec:protocol}

The benchmark follows the initial-structure-to-relaxed-energy protocol of \mbd. Each model is given the unrelaxed WBM structure, that is, the geometry produced by the substitution procedure before any DFT relaxation, and performs a full relaxation of atomic positions and cell. Relaxations use the FIRE optimizer~\cite{bitzek2006fire} with the Fr\'echet cell filter, as implemented in ASE~\cite{larsen2017ase}, a force threshold of 0.02~eV~\AA$^{-1}$ and a maximum of 500 steps. Structures that do not reach the threshold are excluded from the accuracy metrics and reported separately through the convergence rate. The DFT-relaxed structure is used only as the reference for the structural comparison.

For each relaxed structure the predicted total energy is converted into a distance to the reference hull, and stability is defined by a threshold of zero. We report the F1 score, the discovery acceleration factor (\daf), precision, recall, accuracy, the mean absolute error and root-mean-square error of the hull distance, the coefficient of determination $R^2$, and the area under the precision-recall curve. Structural accuracy is measured by the root-mean-square displacement (RMSD) between the relaxed and the DFT geometry. It is computed with \texttt{StructureMatcher}, with a site tolerance of 1.0 and without volume rescaling. Following the convention of \mbd, structures for which no match is found are assigned the value of the tolerance. This fill dominates the reported RMSD, since the matched structures alone have displacements of 0.02 to 0.03~\AA. It nevertheless ranks the models in the same order, and the unmatched structures are precisely those in which the model relaxed out of the DFT basin.

Stability is assessed primarily with the phase separation energy \eps~\cite{bartel2020}, the distance to the hull formed by all compositions other than that of the compound under consideration, for which stable compounds satisfy $\eps\le 0$. The reason is the following. A compound that lies on the hull has $\ehull=0$ exactly, so that any approximate estimate falls on either side of the threshold with roughly equal probability, irrespective of the accuracy of the model. This is irrelevant for the Materials-Project-derived hull, from which the test structures have been removed. It is however important for the deeper \alex\ hull, on which a number of WBM structures do lie, since they are part of \alex. We nevertheless also report \ehull\ against the Materials-Project-derived hull for consistency with \mbd. The two criteria rank the models identically.

The timings of \cref{tab:benchmark} were obtained on a single NVIDIA H200 GPU with one CPU core. The single-point timing is the mean time per structure for the evaluation of the energy, forces and stress of \num{2000} structures of the test set, averaged over three runs after a warm-up run. The molecular-dynamics timing is the mean wall time per step of a microcanonical trajectory of a rattled 1000-atom \ce{MgO} supercell at 300~K with a time step of 2~fs, averaged over 100 steps after 50 warm-up steps. Initialization of the calculator is excluded in both cases. All \tuga\ results in this work, timings included, were obtained with the accurate precision setting of its calculator rather than the reduced-precision default.

\subsection{Models benchmarked}
\label{sec:models}

Only models trained, or fine-tuned, on \matpes-PBE data can be evaluated on this benchmark, since any other uMLIP predicts total energies at a different level of theory. This restricts the set of publicly available candidates considerably. We benchmark the following five models.

\begin{itemize}
  \item \textbf{TensorNet-PES-MatPES-PBE-2025.2}, a Cartesian-tensor equivariant architecture~\cite{simeon2023tensornet}, trained on \matpes-PBE and distributed through MatGL~\cite{ko2025matgl}.
  \item \textbf{M3GNet-PES-MatPES-PBE-2025.2}, a three-body graph network~\cite{chen2022m3gnet}, trained on \matpes-PBE and distributed through MatGL.
  \item \textbf{CHGNet-PES-MatPES-PBE-2025.2.10}, a charge-informed graph network~\cite{deng2023chgnet}, trained on \matpes-PBE and distributed through MatGL.
  \item \textbf{QET-PES-MatPES-PBE-2025.2}, an architecture derived from TensorNet, trained on \matpes-PBE and distributed through MatGL~\cite{ko2025matgl}.
  \item \textbf{MACE-MATPES-PBE-0}, an equivariant message-passing network~\cite{batatia2022mace,batatia2024foundation}. In contrast to the four models above, this one was not trained on \matpes\ alone. It was pretrained on OMat24~\cite{omat24} and subsequently fine-tuned on \matpes-PBE.
\end{itemize}

For the four MatGL models we used the weights retrained after the correction of a bug in the message-passing convention of TensorNet, M3GNet and QET. Weights released before that correction are not numerically valid. All models were evaluated with their released weights, without any further fitting. Every checkpoint was verified before the benchmark, in two ways. We compared its single-point energies on a sample of test structures with the DFT labels, which detects a checkpoint trained at a different level of theory. We also compared its stress with a finite-difference derivative of the energy with respect to the volume, which detects an error in the units of the stress. Two variants of our own model, \tuga, were introduced above in \cref{sec:tuga}.

\subsection{The pressure benchmark}
\label{sec:pressure_methods}

The underlying calculations are those of Ref.~\cite{loew2026pressure}, recomputed with the settings of \cref{sec:dft}. We used the same grid of six pressures, 25, 50, 75, 100, 125 and 150~GPa, with the target pressure imposed through the stress tensor during the relaxation. As for the other datasets used here, a fraction of the structures did not converge within our computational budget, so that the recomputed set is somewhat smaller than the original. The relevant thermodynamic potential is the enthalpy $H=E+PV$, which is the quantity stored and compared throughout. From this set we retain the compounds that were held out from the training of \tuga\ and for which a converged calculation exists at the pressure considered. This yields between \num{7473} and \num{8173} structures per pressure, and 48~thousand in total.

The task is defined in the same way as in \cref{sec:protocol}. Each DFT relaxation at a given pressure was itself started from a structure relaxed at that pressure with GRACE~\cite{lysogorskiy2026grace}, in the course of the high-pressure structure search of Ref.~\cite{pires2026pressure}. It is this structure that is given to the models. They therefore receive exactly the input that the reference calculation received, and must reproduce its result. GRACE is trained at Materials Project settings and is therefore not itself eligible for this benchmark. However, this does not affect its use here. The geometries enter only as the common starting point of the reference calculation and of the models. The baseline they define is purely geometric, and therefore independent of the level of theory at which they were obtained. The relaxations use the same optimizer and convergence criteria as at ambient pressure, with the pressure entering through the cell filter and the $PV$ term added to the energy, so that the predicted quantity is an enthalpy. We verified at ambient pressure that starting a relaxation from the unrelaxed or from the relaxed structure gives indistinguishable results for conservative potentials. The proximity of the starting point to the final one does not by itself make the task easier.

The hulls available at finite pressure are of uneven quality and depart from the construction adopted for the ambient case, so that a stability classification built on them would not be comparable with the one of \cref{sec:ambient}. We therefore restrict the analysis to the quantities that are unambiguously defined, the enthalpy and the relaxed geometry. For each pressure we report the mean and median absolute error of the enthalpy per atom, the mean absolute relative error of the volume per atom, the fraction of relaxations in error by more than 1~\evatom, and the RMSD with respect to the DFT geometry. Since $H=E+PV$, the error of the enthalpy decomposes as $\Delta H=\Delta E+P\,\Delta V$. The two terms distinguish a model that places the cell at the wrong volume from one that assigns the wrong energy to a geometry that is essentially correct. At 150~GPa the $PV$ term amounts to about 11~\evatom, so that a volume error of a few percent is worth several hundred \mevatom\ on its own.

Relaxations that do not reach the force threshold within the step budget are excluded from the mean errors and from the RMSD, as at ambient pressure, and the converged fraction is reported separately. This fraction falls appreciably with pressure for several models. The median error and the fraction of catastrophic failures are therefore computed over all relaxations that returned an energy, converged or not, and the mean over all relaxations is quoted in the text wherever it changes the picture.

\section*{Data availability}

The datasets constructed for this work are released together with the paper at \url{https://alexandria.icams.rub.de/datasets.html}. They comprise four parts. The first is the \wbmmatpes\ test set of 182~thousand structures, with the unrelaxed geometry, the relaxed geometry and the DFT energy of each entry. The second is the Materials Project reference set of 89~thousand structures recomputed at \matpes\ settings, together with the derived convex hull. The third is the \alex\ convex hull at the same settings. The fourth consists of the six pressure test sets, with the starting geometry, the DFT-relaxed geometry and the DFT enthalpy of each entry. All files are lists of \texttt{pymatgen} \texttt{ComputedStructureEntry} objects in compressed JSON format, and no energy corrections have been applied to any of them. The per-structure predictions of the benchmarked models are provided in the same repository. The \alexmatpes\ data used to train \tuga\ will be made available as part of the next release of the \alex\ database.

\section*{Code availability}

The \tuga\ code, together with the trained models used in this work, is available at \url{github.com/ppdebreuck/tuga-ff}.

\section*{Acknowledgements}
T.H.B.d.S. and M.A.L.M. were supported by a collaboration between the Kavli Foundation, Klaus Tschira Stiftung, and Kevin Wells, as part of the SuperC collaboration, and by the Simons Foundation through the Collaboration on New Frontiers in Superconductivity (Grant No.~SFI-MPS-NFS-00006741-10). S.B. acknowledges funding from the Volkswagen Stiftung (Momentum) through the project ``dandelion''. The authors gratefully acknowledge the computing time made available to them on the high-performance computers Noctua and Otus at the NHR Center Paderborn Center for Parallel Computing (PC2). The authors gratefully acknowledge the scientific support and HPC resources provided by the Erlangen National High Performance Computing Center (NHR@FAU) of the Friedrich-Alexander-Universit\"at Erlangen-N\"urnberg (FAU) under the NHR project k114eb. Part of the calculations were performed on the HPC cluster Elysium of the Ruhr University Bochum, subsidized by the DFG (INST 213/1055-1).

\section*{Author contributions}
P.P.D.B. developed the \tuga\ model and carried out its training, with contributions from T.H.B.d.S. A.L. contributed to the development of the architecture and of the training strategy. T.H.B.d.S. and M.A.L.M. constructed the benchmark and performed the density-functional and interatomic-potential calculations. S.B. and M.A.L.M. supervised the work. All authors analyzed the results and contributed to the writing of the manuscript.

\section*{Competing interests}
The authors declare no competing interests.

\bibliography{bib}

@article{bartel2020,
 author = {Bartel, Christopher J. and Trewartha, Amalie and Wang, Qi and Dunn, Alexander and Jain, Anubhav and Ceder, Gerbrand},
 doi = {10.1038/s41524-020-00362-y},
 pages = {97},
 issn = {2057-3960},
 journal = {npj Comput. Mater.},
 month = {July},
 number = {1},
 publisher = {Springer Science and Business Media LLC},
 title = {A critical examination of compound stability predictions from machine-learned formation energies},
 url = {http://dx.doi.org/10.1038/s41524-020-00362-y},
 volume = {6},
 year = {2020}
}

@inproceedings{batatia2022mace,
 author = {Batatia, Ilyes and Kovács, Dávid P. and Simm, Gregor and Ortner, Christoph and Csányi, Gábor},
 booktitle = {Advances in Neural Information Processing Systems 35},
 doi = {10.52202/068431-0830},
 pages = {11423--11436},
 publisher = {Curran Associates, Inc.},
 title = {{MACE}: Higher Order Equivariant Message Passing Neural Networks for Fast and Accurate Force Fields},
 url = {http://dx.doi.org/10.52202/068431-0830},
 year = {2022}
}

@article{batatia2024foundation,
 author = {Batatia, Ilyes and others},
 doi = {10.1063/5.0297006},
 issn = {1089-7690},
 journal = {J. Chem. Phys.},
 number = {18},
 pages = {184110},
 title = {A foundation model for atomistic materials chemistry},
 url = {http://dx.doi.org/10.1063/5.0297006},
 volume = {163},
 year = {2025}
}

@article{bitzek2006fire,
 author = {Bitzek, Erik and Koskinen, Pekka and Gähler, Franz and Moseler, Michael and Gumbsch, Peter},
 doi = {10.1103/physrevlett.97.170201},
 issn = {1079-7114},
 journal = {Phys. Rev. Lett.},
 month = {October},
 number = {17},
 pages = {170201},
 publisher = {American Physical Society (APS)},
 title = {Structural Relaxation Made Simple},
 url = {http://dx.doi.org/10.1103/PhysRevLett.97.170201},
 volume = {97},
 year = {2006}
}

@article{Blchl1994,
 author = {Blöchl, P. E.},
 doi = {10.1103/physrevb.50.17953},
 issn = {1095-3795},
 journal = {Phys. Rev. B},
 month = {December},
 number = {24},
 pages = {17953},
 publisher = {American Physical Society (APS)},
 title = {Projector augmented-wave method},
 url = {http://dx.doi.org/10.1103/PhysRevB.50.17953},
 volume = {50},
 year = {1994}
}

@article{cavignac2026alexandria,
 author = {Cavignac, Théo and Schmidt, Jonathan and De Breuck, Pierre-Paul and Loew, Antoine and Cerqueira, Tiago F T and Wang, Hai-Chen and Bochkarev, Anton and Lysogorskiy, Yury and Romero, Aldo H and Drautz, Ralf and Botti, Silvana and Marques, Miguel A L},
 doi = {10.1088/2515-7639/ae6620},
 issn = {2515-7639},
 journal = {J. Phys.: Mater.},
 month = {May},
 number = {2},
 pages = {025014},
 publisher = {IOP Publishing},
 title = {{AI}-Driven expansion and application of the {Alexandria} database},
 url = {http://dx.doi.org/10.1088/2515-7639/ae6620},
 volume = {9},
 year = {2026}
}

@article{cerqueira2026representation,
 author = {Cerqueira, Tiago F. T. and Wang, Hai-Chen and Botti, Silvana and Marques, Miguel A. L.},
 doi = {10.1038/s41524-025-01916-8},
 pages = {48},
 issn = {2057-3960},
 journal = {npj Comput. Mater.},
 month = {January},
 number = {1},
 publisher = {Springer Science and Business Media LLC},
 title = {A non-orthogonal representation for materials based on chemical similarity},
 url = {http://dx.doi.org/10.1038/s41524-025-01916-8},
 volume = {12},
 year = {2026}
}

@article{chen2022m3gnet,
 author = {Chen, Chi and Ong, Shyue Ping},
 doi = {10.1038/s43588-022-00349-3},
 issn = {2662-8457},
 journal = {Nat. Comput. Sci.},
 month = {November},
 number = {11},
 pages = {718--728},
 publisher = {Springer Science and Business Media LLC},
 title = {A universal graph deep learning interatomic potential for the periodic table},
 url = {http://dx.doi.org/10.1038/s43588-022-00349-3},
 volume = {2},
 year = {2022}
}

@article{deng2023chgnet,
 author = {Deng, Bowen and Zhong, Peichen and Jun, KyuJung and Riebesell, Janosh and Han, Kevin and Bartel, Christopher J. and Ceder, Gerbrand},
 doi = {10.1038/s42256-023-00716-3},
 issn = {2522-5839},
 journal = {Nat. Mach. Intell.},
 month = {September},
 number = {9},
 pages = {1031--1041},
 publisher = {Springer Science and Business Media LLC},
 title = {{CHGNet} as a pretrained universal neural network potential for charge-informed atomistic modelling},
 url = {http://dx.doi.org/10.1038/s42256-023-00716-3},
 volume = {5},
 year = {2023}
}

@article{deng2025softening,
 author = {Deng, Bowen and Choi, Yunyeong and Zhong, Peichen and Riebesell, Janosh and Anand, Shashwat and Li, Zhuohan and Jun, KyuJung and Persson, Kristin A. and Ceder, Gerbrand},
 doi = {10.1038/s41524-024-01500-6},
 pages = {9},
 issn = {2057-3960},
 journal = {npj Comput. Mater.},
 month = {January},
 number = {1},
 publisher = {Springer Science and Business Media LLC},
 title = {Systematic softening in universal machine learning interatomic potentials},
 url = {http://dx.doi.org/10.1038/s41524-024-01500-6},
 volume = {11},
 year = {2025}
}

@misc{dynamat2026,
 author = {Gawkowski, Mikołaj J. and Artrith, Nongnuch and Bonfanti, Silvia and Gangan, Abhijeet Sadashiv and Heenen, Hendrik H. and Kioseoglou, Joseph and Lončarić, Ivor and Myneni, Hemanadhan and Riebesell, Janosh and Rossi, Mariana and Rupp, Matthias and Schmidt, Jonathan and Sharma, Shubham and Shi, Benjamin X. and Wadowski, Antoni and Hörmann, Lukas and Kapil, Venkat},
 copyright = {Creative Commons Attribution Share Alike 4.0 International},
 doi = {10.48550/ARXIV.2607.03433},
 archiveprefix = {arXiv},
 eprint = {2607.03433},
 publisher = {arXiv},
 title = {{Dyna-Mat}: End-to-end benchmarking of foundation machine learning interatomic potentials in finite-temperature ensembles},
 url = {https://arxiv.org/abs/2607.03433},
 year = {2026}
}

@article{huber1964,
 author = {Huber, Peter J.},
 doi = {10.1214/aoms/1177703732},
 issn = {0003-4851},
 journal = {Ann. Math. Stat.},
 month = {March},
 number = {1},
 pages = {73--101},
 publisher = {Institute of Mathematical Statistics},
 title = {Robust Estimation of a Location Parameter},
 url = {http://dx.doi.org/10.1214/aoms/1177703732},
 volume = {35},
 year = {1964}
}

@article{jain2011mixing,
 author = {Jain, Anubhav and Hautier, Geoffroy and Ong, Shyue Ping and Moore, Charles J. and Fischer, Christopher C. and Persson, Kristin A. and Ceder, Gerbrand},
 doi = {10.1103/physrevb.84.045115},
 issn = {1550-235X},
 journal = {Phys. Rev. B},
 month = {July},
 number = {4},
 pages = {045115},
 publisher = {American Physical Society (APS)},
 title = {Formation enthalpies by mixing {GGA} and {GGA+$U$} calculations},
 url = {http://dx.doi.org/10.1103/PhysRevB.84.045115},
 volume = {84},
 year = {2011}
}

@article{jain2013mp,
 author = {Jain, Anubhav and Ong, Shyue Ping and Hautier, Geoffroy and Chen, Wei and Richards, William Davidson and Dacek, Stephen and Cholia, Shreyas and Gunter, Dan and Skinner, David and Ceder, Gerbrand and Persson, Kristin A.},
 doi = {10.1063/1.4812323},
 pages = {011002},
 issn = {2166-532X},
 journal = {APL Mater.},
 month = {July},
 number = {1},
 publisher = {AIP Publishing},
 title = {Commentary: {The Materials Project}: A materials genome approach to accelerating materials innovation},
 url = {http://dx.doi.org/10.1063/1.4812323},
 volume = {1},
 year = {2013}
}

@misc{kingma2015adam,
 author = {Kingma, Diederik P. and Ba, Jimmy},
 copyright = {arXiv.org perpetual, non-exclusive license},
 doi = {10.48550/ARXIV.1412.6980},
 archiveprefix = {arXiv},
 eprint = {1412.6980},
 publisher = {arXiv},
 title = {{Adam}: A Method for Stochastic Optimization},
 url = {https://arxiv.org/abs/1412.6980},
 year = {2014}
}

@article{ko2025matgl,
 author = {Ko, Tsz Wai and Deng, Bowen and Nassar, Marcel and Barroso-Luque, Luis and Liu, Runze and Qi, Ji and Thakur, Atul C. and Mishra, Adesh Rohan and Liu, Elliott and Ceder, Gerbrand and Miret, Santiago and Ong, Shyue Ping},
 doi = {10.1038/s41524-025-01742-y},
 pages = {253},
 issn = {2057-3960},
 journal = {npj Comput. Mater.},
 month = {August},
 number = {1},
 publisher = {Springer Science and Business Media LLC},
 title = {Materials Graph Library ({MatGL}), an open-source graph deep learning library for materials science and chemistry},
 url = {http://dx.doi.org/10.1038/s41524-025-01742-y},
 volume = {11},
 year = {2025}
}

@article{Kresse1996,
 author = {Kresse, G. and Furthmüller, J.},
 doi = {10.1103/physrevb.54.11169},
 issn = {1095-3795},
 journal = {Phys. Rev. B},
 month = {October},
 number = {16},
 pages = {11169},
 publisher = {American Physical Society (APS)},
 title = {Efficient iterative schemes for \textit{ab initio} total-energy calculations using a plane-wave basis set},
 url = {http://dx.doi.org/10.1103/PhysRevB.54.11169},
 volume = {54},
 year = {1996}
}

@article{Kresse1996_2,
 author = {Kresse, G. and Furthmüller, J.},
 doi = {10.1016/0927-0256(96)00008-0},
 issn = {0927-0256},
 journal = {Comp. Mater. Sci.},
 month = {July},
 number = {1},
 pages = {15--50},
 publisher = {Elsevier BV},
 title = {Efficiency of ab-initio total energy calculations for metals and semiconductors using a plane-wave basis set},
 url = {http://dx.doi.org/10.1016/0927-0256(96)00008-0},
 volume = {6},
 year = {1996}
}

@article{Kresse1999,
 author = {Kresse, G. and Joubert, D.},
 doi = {10.1103/physrevb.59.1758},
 issn = {1095-3795},
 journal = {Phys. Rev. B},
 month = {January},
 number = {3},
 pages = {1758},
 publisher = {American Physical Society (APS)},
 title = {From ultrasoft pseudopotentials to the projector augmented-wave method},
 url = {http://dx.doi.org/10.1103/PhysRevB.59.1758},
 volume = {59},
 year = {1999}
}

@article{larsen2017ase,
 author = {Hjorth Larsen, Ask and Jørgen Mortensen, Jens and Blomqvist, Jakob and Castelli, Ivano E and Christensen, Rune and Dułak, Marcin and Friis, Jesper and Groves, Michael N and Hammer, Bjørk and Hargus, Cory and Hermes, Eric D and Jennings, Paul C and Bjerre Jensen, Peter and Kermode, James and Kitchin, John R and Leonhard Kolsbjerg, Esben and Kubal, Joseph and Kaasbjerg, Kristen and Lysgaard, Steen and Bergmann Maronsson, Jón and Maxson, Tristan and Olsen, Thomas and Pastewka, Lars and Peterson, Andrew and Rostgaard, Carsten and Schiøtz, Jakob and Schütt, Ole and Strange, Mikkel and Thygesen, Kristian S and Vegge, Tejs and Vilhelmsen, Lasse and Walter, Michael and Zeng, Zhenhua and Jacobsen, Karsten W},
 doi = {10.1088/1361-648x/aa680e},
 issn = {1361-648X},
 journal = {J. Phys.: Condens. Matter},
 month = {June},
 number = {27},
 pages = {273002},
 publisher = {IOP Publishing},
 title = {The atomic simulation environment---a {Python} library for working with atoms},
 url = {http://dx.doi.org/10.1088/1361-648X/aa680e},
 volume = {29},
 year = {2017}
}

@article{loew2026pressure,
 author = {Loew, Antoine and Schmidt, Jonathan and Botti, Silvana and Marques, Miguel A L},
 doi = {10.1088/2515-7639/ae2ba8},
 issn = {2515-7639},
 journal = {J. Phys.: Mater.},
 month = {December},
 number = {1},
 pages = {015010},
 publisher = {IOP Publishing},
 title = {Universal machine learning potentials under pressure},
 url = {http://dx.doi.org/10.1088/2515-7639/ae2ba8},
 volume = {9},
 year = {2025}
}

@article{lysogorskiy2026grace,
 author = {Lysogorskiy, Yury and Bochkarev, Anton and Drautz, Ralf},
 doi = {10.1038/s41524-026-01979-1},
 pages = {114},
 issn = {2057-3960},
 journal = {npj Comput. Mater.},
 month = {February},
 number = {1},
 publisher = {Springer Science and Business Media LLC},
 title = {Graph atomic cluster expansion for foundational machine learning interatomic potentials},
 url = {http://dx.doi.org/10.1038/s41524-026-01979-1},
 volume = {12},
 year = {2026}
}

@misc{MATPES,
 author = {Kaplan, Aaron D. and Liu, Runze and Qi, Ji and Ko, Tsz Wai and Deng, Bowen and Riebesell, Janosh and Ceder, Gerbrand and Persson, Kristin A. and Ong, Shyue Ping},
 copyright = {arXiv.org perpetual, non-exclusive license},
 doi = {10.48550/ARXIV.2503.04070},
 archiveprefix = {arXiv},
 eprint = {2503.04070},
 publisher = {arXiv},
 title = {A Foundational Potential Energy Surface Dataset for Materials},
 url = {https://arxiv.org/abs/2503.04070},
 year = {2025}
}

@article{merchant2023gnome,
 author = {Merchant, Amil and Batzner, Simon and Schoenholz, Samuel S. and Aykol, Muratahan and Cheon, Gowoon and Cubuk, Ekin Dogus},
 doi = {10.1038/s41586-023-06735-9},
 issn = {1476-4687},
 journal = {Nature},
 month = {November},
 number = {7990},
 pages = {80--85},
 publisher = {Springer Science and Business Media LLC},
 title = {Scaling deep learning for materials discovery},
 url = {http://dx.doi.org/10.1038/s41586-023-06735-9},
 volume = {624},
 year = {2023}
}

@article{mpaloe2025,
 author = {Kuner, Matthew C. and Kaplan, Aaron D. and Persson, Kristin A. and Asta, Mark and Chrzan, Daryl C.},
 doi = {10.1038/s41524-025-01834-9},
 pages = {352},
 issn = {2057-3960},
 journal = {npj Comput. Mater.},
 month = {November},
 number = {1},
 publisher = {Springer Science and Business Media LLC},
 title = {{MP-ALOE}: an {r$^2$SCAN} dataset for universal machine learning interatomic potentials},
 url = {http://dx.doi.org/10.1038/s41524-025-01834-9},
 volume = {11},
 year = {2025}
}

@misc{neumann2024orb,
 author = {Neumann, Mark and Gin, James and Rhodes, Benjamin and Bennett, Steven and Li, Zhiyi and Choubisa, Hitarth and Hussey, Arthur and Godwin, Jonathan},
 copyright = {Creative Commons Attribution 4.0 International},
 doi = {10.48550/ARXIV.2410.22570},
 archiveprefix = {arXiv},
 eprint = {2410.22570},
 publisher = {arXiv},
 title = {{Orb}: A Fast, Scalable Neural Network Potential},
 url = {https://arxiv.org/abs/2410.22570},
 year = {2024}
}

@misc{omat24,
 author = {Barroso-Luque, Luis and Shuaibi, Muhammed and Fu, Xiang and Wood, Brandon M. and Dzamba, Misko and Gao, Meng and Rizvi, Ammar and Zitnick, C. Lawrence and Ulissi, Zachary W.},
 copyright = {arXiv.org perpetual, non-exclusive license},
 doi = {10.48550/ARXIV.2410.12771},
 archiveprefix = {arXiv},
 eprint = {2410.12771},
 publisher = {arXiv},
 title = {{Open Materials 2024 (OMat24)} Inorganic Materials Dataset and Models},
 url = {https://arxiv.org/abs/2410.12771},
 year = {2024}
}

@article{ong2013pymatgen,
 author = {Ong, Shyue Ping and Richards, William Davidson and Jain, Anubhav and Hautier, Geoffroy and Kocher, Michael and Cholia, Shreyas and Gunter, Dan and Chevrier, Vincent L. and Persson, Kristin A. and Ceder, Gerbrand},
 doi = {10.1016/j.commatsci.2012.10.028},
 issn = {0927-0256},
 journal = {Comp. Mater. Sci.},
 month = {February},
 pages = {314--319},
 publisher = {Elsevier BV},
 title = {{Python Materials Genomics (pymatgen)}: A robust, open-source {Python} library for materials analysis},
 url = {http://dx.doi.org/10.1016/j.commatsci.2012.10.028},
 volume = {68},
 year = {2013}
}

@article{park2024sevennet,
 author = {Park, Yutack and Kim, Jaesun and Hwang, Seungwoo and Han, Seungwu},
 doi = {10.1021/acs.jctc.4c00190},
 issn = {1549-9626},
 journal = {J. Chem. Theory Comput.},
 month = {May},
 number = {11},
 pages = {4857--4868},
 publisher = {American Chemical Society (ACS)},
 title = {Scalable Parallel Algorithm for Graph Neural Network Interatomic Potentials in Molecular Dynamics Simulations},
 url = {http://dx.doi.org/10.1021/acs.jctc.4c00190},
 volume = {20},
 year = {2024}
}

@article{Perdew1996,
 author = {Perdew, John P. and Burke, Kieron and Ernzerhof, Matthias},
 doi = {10.1103/physrevlett.77.3865},
 issn = {1079-7114},
 journal = {Phys. Rev. Lett.},
 month = {October},
 number = {18},
 pages = {3865},
 publisher = {American Physical Society (APS)},
 title = {Generalized Gradient Approximation Made Simple},
 url = {http://dx.doi.org/10.1103/PhysRevLett.77.3865},
 volume = {77},
 year = {1996}
}

@unpublished{pires2026pressure,
 author = {Pires, Paulo R. and De Breuck, Pierre-Paul and Wang, Hai-Chen and
Cavignac, Th\'{e}o and Bochkarev, Anton and Lysogorskiy, Yury and
Drautz, Ralf and Shi, Jingming and Botti, Silvana and
Marques, Miguel A. L.},
 note = {Submitted},
 title = {Machine learning accelerated exploration of the global high-pressure phase diagram},
 year = {2026}
}

@article{Riebesell2025,
 author = {Riebesell, Janosh and Goodall, Rhys E. A. and Benner, Philipp and Chiang, Yuan and Deng, Bowen and Ceder, Gerbrand and Asta, Mark and Lee, Alpha A. and Jain, Anubhav and Persson, Kristin A.},
 doi = {10.1038/s42256-025-01055-1},
 issn = {2522-5839},
 journal = {Nat. Mach. Intell.},
 month = {June},
 number = {6},
 pages = {836--847},
 publisher = {Springer Science and Business Media LLC},
 title = {A framework to evaluate machine learning crystal stability predictions},
 url = {http://dx.doi.org/10.1038/s42256-025-01055-1},
 volume = {7},
 year = {2025}
}

@article{schmidt2022alexandria,
 author = {Schmidt, Jonathan and Wang, Hai-Chen and Cerqueira, Tiago F. T. and Botti, Silvana and Marques, Miguel A. L.},
 doi = {10.1038/s41597-022-01177-w},
 pages = {64},
 issn = {2052-4463},
 journal = {Sci. Data},
 month = {March},
 number = {1},
 publisher = {Springer Science and Business Media LLC},
 title = {A dataset of 175k stable and metastable materials calculated with the {PBEsol} and {SCAN} functionals},
 url = {http://dx.doi.org/10.1038/s41597-022-01177-w},
 volume = {9},
 year = {2022}
}

@article{schmidt2023alexandria,
 author = {Schmidt, Jonathan and Hoffmann, Noah and Wang, Hai-Chen and Borlido, Pedro and Carriço, Pedro J. M. A. and Cerqueira, Tiago F. T. and Botti, Silvana and Marques, Miguel A. L.},
 doi = {10.1002/adma.202210788},
 pages = {2210788},
 issn = {1521-4095},
 journal = {Adv. Mater.},
 month = {April},
 number = {22},
 publisher = {Wiley},
 title = {Machine-Learning-Assisted Determination of the Global Zero-Temperature Phase Diagram of Materials},
 url = {http://dx.doi.org/10.1002/adma.202210788},
 volume = {35},
 year = {2023}
}

@article{schmidt2024alexandria,
 author = {Schmidt, Jonathan and Cerqueira, Tiago F. T. and Romero, Aldo H. and Loew, Antoine and Jäger, Fabian and Wang, Hai-Chen and Botti, Silvana and Marques, Miguel A. L.},
 doi = {10.1016/j.mtphys.2024.101560},
 issn = {2542-5293},
 journal = {Mater. Today Phys.},
 month = {November},
 pages = {101560},
 publisher = {Elsevier BV},
 title = {Improving machine-learning models in materials science through large datasets},
 url = {http://dx.doi.org/10.1016/j.mtphys.2024.101560},
 volume = {48},
 year = {2024}
}

@misc{shazeer2020glu,
 author = {Shazeer, Noam},
 copyright = {arXiv.org perpetual, non-exclusive license},
 doi = {10.48550/ARXIV.2002.05202},
 archiveprefix = {arXiv},
 eprint = {2002.05202},
 publisher = {arXiv},
 title = {{GLU} Variants Improve Transformer},
 url = {https://arxiv.org/abs/2002.05202},
 year = {2020}
}

@inproceedings{simeon2023tensornet,
 author = {Simeon, Guillem and De Fabritiis, Gianni},
 booktitle = {Advances in Neural Information Processing Systems 36},
 doi = {10.52202/075280-1623},
 pages = {37334--37353},
 publisher = {Curran Associates, Inc.},
 title = {{TensorNet}: {Cartesian} Tensor Representations for Efficient Learning of Molecular Potentials},
 url = {http://dx.doi.org/10.52202/075280-1623},
 year = {2023}
}

@inproceedings{smith2019onecycle,
 author = {Smith, Leslie N. and Topin, Nicholay},
 booktitle = {Artificial Intelligence and Machine Learning for Multi-Domain Operations Applications},
 doi = {10.1117/12.2520589},
 editor = {Pham, Tien},
 month = {May},
 pages = {1100612},
 publisher = {SPIE},
 series = {Proc. SPIE},
 volume = {11006},
 title = {Super-convergence: very fast training of neural networks using large learning rates},
 url = {http://dx.doi.org/10.1117/12.2520589},
 year = {2019}
}

@article{wang2006ggau,
 author = {Wang, Lei and Maxisch, Thomas and Ceder, Gerbrand},
 doi = {10.1103/physrevb.73.195107},
 issn = {1550-235X},
 journal = {Phys. Rev. B},
 month = {May},
 number = {19},
 pages = {195107},
 publisher = {American Physical Society (APS)},
 title = {Oxidation energies of transition metal oxides within the {GGA+$U$} framework},
 url = {http://dx.doi.org/10.1103/PhysRevB.73.195107},
 volume = {73},
 year = {2006}
}

@article{wang2021mp2020,
 author = {Wang, Amanda and Kingsbury, Ryan and McDermott, Matthew and Horton, Matthew and Jain, Anubhav and Ong, Shyue Ping and Dwaraknath, Shyam and Persson, Kristin A.},
 doi = {10.1038/s41598-021-94550-5},
 pages = {15496},
 issn = {2045-2322},
 journal = {Sci. Rep.},
 month = {July},
 number = {1},
 publisher = {Springer Science and Business Media LLC},
 title = {A framework for quantifying uncertainty in {DFT} energy corrections},
 url = {http://dx.doi.org/10.1038/s41598-021-94550-5},
 volume = {11},
 year = {2021}
}

@article{wang2021wbm,
 author = {Wang, Hai-Chen and Botti, Silvana and Marques, Miguel A. L.},
 doi = {10.1038/s41524-020-00481-6},
 pages = {12},
 issn = {2057-3960},
 journal = {npj Comput. Mater.},
 month = {January},
 number = {1},
 publisher = {Springer Science and Business Media LLC},
 title = {Predicting stable crystalline compounds using chemical similarity},
 url = {http://dx.doi.org/10.1038/s41524-020-00481-6},
 volume = {7},
 year = {2021}
}

@misc{wood2025uma,
 author = {Wood, Brandon M. and Dzamba, Misko and Fu, Xiang and Gao, Meng and Shuaibi, Muhammed and Barroso-Luque, Luis and Abdelmaqsoud, Kareem and Gharakhanyan, Vahe and Kitchin, John R. and Levine, Daniel S. and Michel, Kyle and Sriram, Anuroop and Cohen, Taco and Das, Abhishek and Rizvi, Ammar and Sahoo, Sushree Jagriti and Ulissi, Zachary W. and Zitnick, C. Lawrence},
 copyright = {arXiv.org perpetual, non-exclusive license},
 doi = {10.48550/ARXIV.2506.23971},
 archiveprefix = {arXiv},
 eprint = {2506.23971},
 publisher = {arXiv},
 title = {{UMA}: A Family of Universal Models for Atoms},
 url = {https://arxiv.org/abs/2506.23971},
 year = {2025}
}

@article{Xia2025,
 author = {Xia, Junfan and Zhang, Yaolong and Jiang, Bin},
 doi = {10.1039/d5cs00104h},
 issn = {1460-4744},
 journal = {Chem. Soc. Rev.},
 number = {10},
 pages = {4790--4821},
 publisher = {Royal Society of Chemistry (RSC)},
 title = {The evolution of machine learning potentials for molecules, reactions and materials},
 url = {http://dx.doi.org/10.1039/d5cs00104h},
 volume = {54},
 year = {2025}
}

@misc{xiong2020prenorm,
 author = {Xiong, Ruibin and Yang, Yunchang and He, Di and Zheng, Kai and Zheng, Shuxin and Xing, Chen and Zhang, Huishuai and Lan, Yanyan and Wang, Liwei and Liu, Tie-Yan},
 copyright = {arXiv.org perpetual, non-exclusive license},
 doi = {10.48550/ARXIV.2002.04745},
 archiveprefix = {arXiv},
 eprint = {2002.04745},
 publisher = {arXiv},
 title = {On Layer Normalization in the Transformer Architecture},
 url = {https://arxiv.org/abs/2002.04745},
 year = {2020}
}

@misc{yang2024mattersim,
 author = {Yang, Han and Hu, Chenxi and Zhou, Yichi and Liu, Xixian and Shi, Yu and Li, Jielan and Li, Guanzhi and Chen, Zekun and Chen, Shuizhou and Zeni, Claudio and Horton, Matthew and Pinsler, Robert and Fowler, Andrew and Zügner, Daniel and Xie, Tian and Smith, Jake and Sun, Lixin and Wang, Qian and Kong, Lingyu and Liu, Chang and Hao, Hongxia and Lu, Ziheng},
 copyright = {arXiv.org perpetual, non-exclusive license},
 doi = {10.48550/ARXIV.2405.04967},
 archiveprefix = {arXiv},
 eprint = {2405.04967},
 publisher = {arXiv},
 title = {{MatterSim}: A Deep Learning Atomistic Model Across Elements, Temperatures and Pressures},
 url = {https://arxiv.org/abs/2405.04967},
 year = {2024}
}

@misc{zhang2019rmsnorm,
 author = {Zhang, Biao and Sennrich, Rico},
 copyright = {arXiv.org perpetual, non-exclusive license},
 doi = {10.48550/ARXIV.1910.07467},
 archiveprefix = {arXiv},
 eprint = {1910.07467},
 publisher = {arXiv},
 title = {Root Mean Square Layer Normalization},
 url = {https://arxiv.org/abs/1910.07467},
 year = {2019}
}

\clearpage
\appendix
\newpage
\onecolumngrid

\section{Architecture details and hyperparameters}
\label{app:architecture}

Architecture and training hyperparameters are compiled in \cref{tab:arch_hparams}.

\subsection{\texorpdfstring{$C^4$}{C4} polynomial cutoff envelope}
\label{app:cutoff_poly}

Interatomic distances $r$ are modulated by a polynomial envelope $\chi(r)$ that decays over a boundary window $\omega = 1.0$~\AA\ and vanishes identically for $r \ge r_c$:
\begin{equation}
  \chi(r) = \begin{cases}
    1, & r < r_c - \omega, \\
    p\Big(\frac{r - (r_c - \omega)}{\omega}\Big), & r_c - \omega \le r \le r_c, \\
    0, & r > r_c,
  \end{cases}
  \label{eq:envelope}
\end{equation}
with normalized coordinate $u = [r - (r_c - \omega)]/\omega \in [0, 1]$ and switching polynomial
\begin{equation}
  p(u) = 1 - 126 u^5 + 420 u^6 - 540 u^7 + 315 u^8 - 70 u^9.
  \label{eq:degree9_poly}
\end{equation}
Because boundary derivatives vanish through fourth order ($p^{(k)}(0) = p^{(k)}(1) = 0$ for $k \in \{1, 2, 3, 4\}$), $\chi(r)$ is $C^4$-continuous, yielding $C^3$ forces and $C^2$ vibrational dynamical matrices across the cutoff boundary.

\subsection{Symmetric update operator}
\label{app:sym_update}

The update operator $\mathrm{SymUpdate}(\mathbf t, \mathbf u, \mathbf w)$ updates target states $\mathbf t \in \mathbb R^{d_t}$ from unordered source pairs $\mathbf u, \mathbf w \in \mathbb R^{d_s}$ (\cref{fig:architecture}c). Source states are encoded by a shared MLP ($\tilde{\mathbf u} = \mathrm{MLP}(\mathbf u)$, $\tilde{\mathbf w} = \mathrm{MLP}(\mathbf w)$) and pooled symmetrically:
\begin{equation}
  \mathbf s = \frac{1}{\sqrt 2}\big(\tilde{\mathbf u} + \tilde{\mathbf w}\big), \qquad
  \mathbf p = \tilde{\mathbf u} \odot \tilde{\mathbf w}.
\end{equation}
The target state is updated through a gated SwiGLU projection:
\begin{align}
  [\mathbf f_v, \mathbf f_g] &= \mathbf W_t\,\mathrm{RMSNorm}(\mathbf t) + \mathbf W_s\,\mathbf s + \mathbf W_p\,\mathbf p, \\
  \mathbf c &= \mathrm{RMSNorm}\Big(\mathbf W_c\,\big(\mathrm{SiLU}(\mathbf f_v) \odot \mathbf f_g\big)\Big), \\
  \mathrm{SymUpdate}(\mathbf t, \mathbf u, \mathbf w) &= 2\,\sigma\big(\mathbf w_m^\top \mathbf c + b_m\big) \cdot \big(\mathbf W_m\,\mathbf c\big). \label{eq:sym_result}
\end{align}
Pre-encoding $\tilde{\mathbf u}$ and $\tilde{\mathbf w}$ on nodes before triplet indexing avoids materializing $(N_{\mathrm{triplets}}, 2d)$ tensors in memory.

\subsection{FiLM-modulated graph attention}
\label{app:film_attn}

\begin{figure*}[p]
  \centering
  \includegraphics[width=\textwidth]{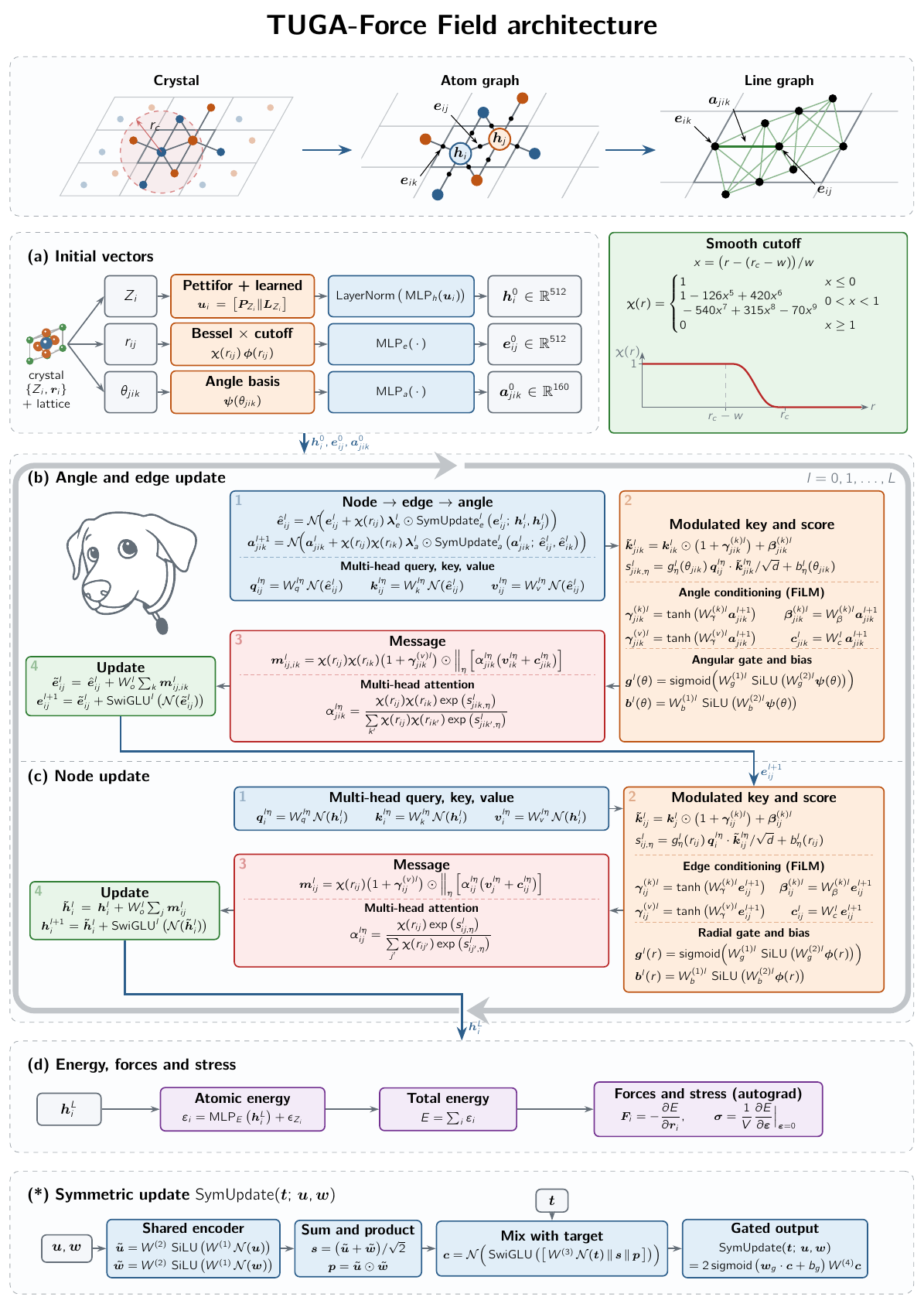}
  \label{fig:architecture}
\end{figure*}

Node states provide queries, keys, and values $\mathbf q_i = \mathbf W_q \mathbf x_i$, $\mathbf k_j = \mathbf W_k \mathbf x_j$, and $\mathbf v_j = \mathbf W_v \mathbf x_j$ (\cref{fig:architecture}b). Edge representations $\mathbf e_{ij}$ modulate keys and values:
\begin{align}
  \tilde{\mathbf k}_{ij} &= \mathbf k_j \odot \big(1 + \tanh\boldsymbol\gamma^{k}_{ij}\big) + \boldsymbol\beta^{k}_{ij}, \\
  \tilde{\mathbf v}_{ij} &= \big(\mathbf v_j + \mathbf e^{v}_{ij}\big)\odot \big(1 + \tanh\boldsymbol\gamma^{v}_{ij}\big),
\end{align}
with $[\mathbf e^{v}_{ij},\,\boldsymbol\gamma^{v}_{ij},\,\boldsymbol\gamma^{k}_{ij},\,\boldsymbol\beta^{k}_{ij}] = \mathbf W_e\,\mathbf e_{ij}$. Attention weights and messages incorporate envelope gating $\chi(r_{ij})$:
\begin{align}
  a_{ij} &= g(r_{ij})\,\frac{\mathbf q_i^\top \tilde{\mathbf k}_{ij}}{\sqrt{d_h}} + b(r_{ij}), \\
  \alpha_{ij} &= \frac{\exp(a_{ij})\,\chi(r_{ij})}{\sum_{k\in\mathcal N(i)}\exp(a_{ik})\,\chi(r_{ik})}, \qquad
  \mathbf m_{ij} = \tilde{\mathbf v}_{ij}\,\alpha_{ij}\,\chi(r_{ij}),
\end{align}
where $g(r_{ij})$ and $b(r_{ij})$ are linear projections of the radial Bessel expansion.

\clearpage
\clearpage

\subsection{Numerical precision and runtime profiles}
\label{app:precision}

Geometric quantities (distances, angles, basis functions, and cutoff envelopes) are evaluated in float32 or float64 before casting to the transformer backbone. Three runtime profiles are supported:
\begin{itemize}\setlength{\itemsep}{0pt}\setlength{\parskip}{0pt}
  \item \texttt{fast}: Geometry and readout in float32, backbone in bfloat16 (TF32 matmuls).
  \item \texttt{balanced}: Full IEEE float32 throughout geometry, backbone, and heads (default).
  \item \texttt{precise}: Full IEEE float64 throughout geometry, backbone, and autograd passes.
\end{itemize}

\begin{table}[h]
  \centering
  \caption{Hyperparameters of the \tuga\ architecture and training protocol. Element-dependent
    cutoff radii cover all 89 elements.}
  \label{tab:arch_hparams}
  \small
  \renewcommand{\arraystretch}{0.92}
  \begin{tabular}{ll@{\hspace{2.2em}}ll}
    \hline\hline
    \multicolumn{2}{l}{\textit{Architecture}} & \multicolumn{2}{l}{\textit{Training and Deployment}} \\[2pt]
    \hline
    Total parameters & \num{44129991} & Loss function & Huber [Eq.~(\ref{eq:huber_loss})] \\
    Model width $d$ & 512 & Loss weights & $(w_E, w_F, w_S) = (2.0, 2.0, 0.4)$ \\
    Interaction layers $L$ & 4 & Energy threshold & $\delta_E = 10$~\mevatom \\
    Attention heads $H$ & 8 ($d_h = 64$) & Force threshold & $\delta_F = 30$~$\mathrm{meV}\,\text{\AA}^{-1}$ \\
    FFN expansion & 4 (SwiGLU, 2048) & Stress threshold & $\delta_S = 4$~$\mathrm{meV}\,\text{\AA}^{-3}$ \\
    Angular channel width & 160 & Optimizer & Adam ($\beta_1=0.9, \beta_2=0.99$) \\
    Update hidden dim. & 160 & Weight decay & $10^{-4}$ \\
    Permutation pooling & SymUpdate & Peak learning rate & $10^{-4}$ (pre), $10^{-5}$ (fine) \\
    Atom embedding dim. & 128 & LR schedule & One-cycle, 5\% warmup \\
    Physical embedding dim. & 98 & Gradient clipping & Norm 0.5 \\
    Distance basis & 64 Bessel ($r_c = 8.0$~\AA) & Batch budget & 240{,}000 atoms \\
    Angular basis & 96 Gaussians ($\cos\theta$) & Species energy shift & MP2020 references \\
    Triplet base cutoff & 4.0~\AA & Pre-training steps & \num{1500000} (OMat24) \\
    Smoothing window $\omega$ & 1.0~\AA & Fine-tuning steps & 50\,000--200\,000 \\
    Bond cutoff radii $R_c$ & 3.76--7.49~\AA & Inference (default) & Balanced (IEEE float32) \\
    Triplet cutoff radii $R_c^\tau$ & 2.90--5.28~\AA & Inference (precise) & Precise (IEEE float64) \\
    Max atom neighbours & 150 & Hardware & NVIDIA H100 SXM5 (80~GB) \\
    Max triplet legs & 50 & Distributed engine & PyTorch Lightning DDP \\
    Cutoff envelope & $C^4$ polynomial [Eq.~(\ref{eq:envelope})] & Pre-train precision & TF32 \\
    \hline\hline
  \end{tabular}
\end{table}

\section{Training protocol and dataset specifications}
\label{app:training}

\subsection{Multi-task Huber loss}
\label{app:loss}

Training minimizes a multi-task Huber loss~\cite{huber1964} on energies, forces, and Cauchy stress tensors:
\begin{equation}
  \mathcal L = w_E\,\rho_{\delta_E}\Big(\frac{E - E^*}{N}\Big)
    + w_F\,\rho_{\delta_F}(\mathbf F - \mathbf F^*)
    + w_S\,\rho_{\delta_S}(\boldsymbol\sigma - \boldsymbol\sigma^*),
  \label{eq:huber_loss}
\end{equation}
where $N$ is the number of atoms, asterisks indicate DFT reference values, and
$\rho_\delta(x) = \frac{1}{2} x^2$ for $|x| \le \delta$ and $\delta (|x| - \frac{1}{2}\delta)$ for $|x| > \delta$.
The loss weights are $w_E = 2.0$, $w_F = 2.0$, and $w_S = 0.4$, with thresholds
$\delta_E = 10$~\mevatom, $\delta_F = 30$~$\mathrm{meV}\,\text{\AA}^{-1}$, and
$\delta_S = 4$~$\mathrm{meV}\,\text{\AA}^{-3}$ ($\approx 0.64$~GPa).

\subsection{Training stages and checkpoints}
\label{app:checkpoints}

Models are pre-trained on OMat24~\cite{omat24} and fine-tuned on targeted corpora:
\begin{itemize}\setlength{\itemsep}{1pt}\setlength{\parskip}{0pt}
  \item \textbf{Pre-training (base):} 1.5~million steps on OMat24 (89.5~million frames from AIMD at 1000 and 3000~K, rattled structures and relaxations). The initial $10^6$ steps at learning rate $10^{-4}$, followed by $5\times 10^5$ steps at $3\times 10^{-5}$.
  \item \texttt{v1-mp}: 200~thousand fine-tuning steps (peak learning rate $2\times 10^{-5}$) on 12~million frames from MPtrj~\cite{deng2023chgnet} and \alex~\cite{schmidt2022alexandria} under Materials Project DFT parameters.
  \item \texttt{v1-matpes-3d}: 50~thousand fine-tuning steps (peak learning rate $10^{-5}$) on the 7~million 3D crystals of \alexmatpes, conforming to \cref{sec:dft}.
  \item \texttt{v1-matpes-all}: 200~thousand fine-tuning steps (peak learning rate $10^{-5}$) on the 21.4~million frames of the full \alexmatpes\ set, across 0D clusters, 1D wires, 2D sheets, 3D bulk, and high-pressure structures up to \qty{150}{\giga\pascal}.
\end{itemize}
WBM test structures and high-pressure benchmark compounds were strictly withheld from training.

\subsection{Computational budget}
\label{app:compute}

Training runs were performed on NVIDIA H100 SXM5 (80~GB) nodes using PyTorch Lightning DDP.
Budgets and allocations are compiled in \cref{tab:train_runs}.

\begin{table}[H]
  \centering
  \caption{Training runs and computational budget for the \tuga\ models.
    Pre-training on OMat24 produces the base model, and the three fine-tunes
    yield the released checkpoints.}
  \label{tab:train_runs}
  \small
  \begin{tabular}{llrcccc}
    \hline\hline
    Run & Dataset & Frames & Steps & Wall time & GPUs & GPU-hours \\
    \hline
    Pre-training & OMat24 & \num{89505421} & \num{1500000} & 200~h & 7\,$\times$\,H100 & \num{1402} \\
    \texttt{v1-mp} & MPtrj + sAlex & \num{12028160} & \num{200000} & 38~h & 7\,$\times$\,H100 & \num{265} \\
    \texttt{v1-matpes-3d} & \alexmatpes\ 3D & \num{7038067} & \num{50000} & 9.3~h & 6\,$\times$\,H100 & \num{56} \\
    \texttt{v1-matpes-all} & \alexmatpes\ all & \num{21379291} & \num{200000} & 38~h & 4\,$\times$\,H100 & \num{152} \\
    \hline\hline
  \end{tabular}
\end{table}

\FloatBarrier
\subsection{Dataset composition}
\label{app:dataset_breakdown}

Structure counts and data-loader sampling weights are compiled in \cref{tab:dataset_counts}.

\begin{table}[!htbp]
  \centering
  \caption{Structure counts and data-loader sampling weights across pre-training and fine-tuning datasets.
    Low-dimensional data includes 20~thousand 0D molecules and clusters in vacuum supercells from \alex.}
  \label{tab:dataset_counts}
  \small
  \begin{tabular}{lrr@{\hspace{2em}}lrr}
    \hline\hline
    Dataset / Split & Frames & Weight & Dataset / Split & Frames & Weight \\
    \hline
    \multicolumn{3}{l}{\textit{OMat24 Pre-training}} & \multicolumn{3}{l}{\textit{\tuga-v1-all Fine-tuning}} \\
    AIMD PBE 1000\,K (NPT) & \num{21269486} & 4.000 & 3D bulk crystals & \num{7038067} & 9.072 \\
    AIMD PBE 1000\,K (NVT) & \num{20256650} & 4.000 & 1D nanowires / chains & \num{8855} & 0.011 \\
    AIMD PBE 3000\,K (NPT) & \num{6076290} & 1.500 & 2D monolayers / sheets & \num{354495} & 0.457 \\
    AIMD PBE 3000\,K (NVT) & \num{7839846} & 1.500 & High pressure (\qty{25}{\giga\pascal}) & \num{1180220} & 1.521 \\
    Rattled 1000\,K & \num{11388510} & 1.000 & High pressure (\qty{50}{\giga\pascal}) & \num{5052760} & 6.513 \\
    Rattled 500\,K & \num{6922197} & 1.000 & High pressure (\qty{75}{\giga\pascal}) & \num{2168452} & 2.795 \\
    Rattled 300\,K & \num{6319139} & 1.000 & High pressure (\qty{100}{\giga\pascal}) & \num{2608755} & 3.363 \\
    Rattled relaxations & \num{9433303} & 2.000 & High pressure (\qty{125}{\giga\pascal}) & \num{1083895} & 1.397 \\
    \textbf{OMat24 subtotal} & \textbf{\num{89505421}} & -- & High pressure (\qty{150}{\giga\pascal}) & \num{1021529} & 1.317 \\
    \multicolumn{3}{l}{\textit{\tuga-v1-mp Fine-tuning}} & Single-point & \num{429074} & 0.553 \\
    Alexandria trajectories (\texttt{salex}) & \num{10447765} & 2.000 & MP-\matpes-PBE (2025.2) & \num{433189} & 2.185 \\
    Materials Project Trajectory (MPtrj) & \num{1580395} & 1.000 & \textbf{\tuga-v1-all subtotal} & \textbf{\num{21379291}} & -- \\
    \textbf{\tuga-v1-mp subtotal} & \textbf{\num{12028160}} & -- & & & \\
    \multicolumn{3}{l}{\textit{\tuga-v1-3d Fine-tuning}} & & & \\
    3D bulk crystals & \num{7038067} & 1.000 & & & \\
    \textbf{\tuga-v1-3d subtotal} & \textbf{\num{7038067}} & -- & & & \\
    \hline\hline
  \end{tabular}
\end{table}

\twocolumngrid

\end{document}